\documentclass[conference]{IEEEtran}
\pdfoutput=1
\usepackage{amsmath,amssymb,amsfonts}
\usepackage{algorithmic}
\usepackage{graphicx, caption, subcaption}
\usepackage{textcomp}
\usepackage{gensymb}
\usepackage{hyperref}

\def\BibTeX{{\rm B\kern-.05em{\sc i\kern-.025em b}\kern-.08em
    T\kern-.1667em\lower.7ex\hbox{E}\kern-.125emX}}

\graphicspath{{../Figures/forPaper/fromRegression/}, {../Figures/forPaper/fromRawData/}}
\title{The COVID-19 pandemic’s impact on U.S.
electricity demand and supply: an early view from the data}
\author{\IEEEauthorblockN{Duzgun Agdas}\IEEEauthorblockA{School of Civil and Environmental Engineering \\ Queensland University of Technology, \\ Brisbane, Queensland 4000, Australia \\ (email: duzgun.agdas@qut.edu.au)}\and
  \IEEEauthorblockN{Prabir Barooah}\IEEEauthorblockA{Department of Mechanical Engineering, \\
    University of Florida, \\
    Gainesville, FL 32611, USA \\
    (email: pbarooah@ufl.edu)}}

\begin{document}
\maketitle

\begin{abstract}
After the onset of the recent COVID-19 pandemic, a number of studies reported on possible changes in electricity consumption trends. The overall theme of these reports was that ``electricity use has decreased during the pandemic, but the power grid is still reliable''.  In this paper we analyze electricity data upto end of May 2020, examining both electricity demand and variables that can indicate stress on the power grid, such as peak demand and demand ramp-rate.  We limit this study to three states in the USA: New York, California, and Florida. The results indicate that the effect of the pandemic on electricity demand is not a simple reduction from comparable time frames, and there are noticeable differences among regions. The variables that can indicate stress on the grid also conveyed mixed messages: some indicate an increase in stress, some indicate a decrease, and some do not indicate any clear difference.  A positive message is that some of the changes that were observed around the time stay-at-home orders were issued appeared to revert back by May 2020.  A key challenge in ascribing any observed change to the pandemic is correcting for weather.  We provide a weather-correction method, apply it to a small city-wide area, and discuss the implications of the estimated changes in demand. The weather correction exercise underscored that weather-correction is as challenging as it is important. 
\end{abstract}


\section{Introduction}
\label{sec:introduction}
Reliable electricity supply is a fundamental service for a functioning society. Since a large part of the workforce is working from home due to the COVID-19 pandemic,  uninterrupted electricity supply has become even more important. The other important service for the nation's productivity during these unprecedented times is internet connectivity,  for which electricity is a prerequisite.  Thus, the potential impact of the COVID-19 pandemic on electricity supply and demand is of interest to many.

Understanding demand changes is important to balancing authorities that are in charge of maintaining reliable operation of the power grid. The New York Independent System Operator (NYISO) has reported large drops in electricity demand after the pandemic, in the range of $10\%$, below typical levels~\cite{nyiso_covid:2020}.  California ISO used a backcasting method to account for weather, and reported a reduction of $0.5-12\%$ in hourly demand in the week after the stay-at-home order~\cite{caiso_covid:May272020}. A reduction of electricity consumption to the tune of 10\% in Europe is reported in~\cite{CicalaEarly:2020}. A number of media reports have drawn from these studies~\cite{BuiAnother:April8_2020,wnn_USeconomists:April172020}.  More recently,  the Energy Information Agency (EIA) reports that while electricity demand seems to have reduced in many parts of the U.S.,  there are variabilities in these trends, and specifically Florida demand trends appear not to indicate a reduction~\cite{eia_florida_covid:20}.

Reliability of the nation's electricity system depends less on electric energy consumption than on variability of demand and generation.  Large changes in demand makes it harder for the operators of the power grids to maintain demand-supply balance,  which is critical for reliability.  So far, the power grids throughout the USA have continued to deliver electricity reliably.  Grid operators took special precautions early on to prevent disruptions to grid operations due to the pandemic~\cite{EPRICOVID_March27:2020,news_NISO:April7_2020}.  Some potential issues such as the challenges of refueling nuclear power plants are being worked out.  In contrast to electricity demand, there is a lack of studies examining the effect of the COVID-19 mitigation measures on the power grid's reliability.

Majority of the studies described above were published soon after the stay-at-home orders were issued.  Now that a few more weeks have passed, it is useful to examine the data again to see if the trends reported in the early studies hold.

In this paper we analyze data from U.S. Energy Information Administration (EIA) until the end of May 2020 to examine trends in electricity demand and indicators of stress on the power grid.  We focus on data from three states in the USA: Florida, New York, and California.  There are few reasons for this short list. One, these three states are similar in the nominal electricity consumption but distinct in both geography, demographics, and prevalence of COVID-19. Two, prior work exist on NY and CA, but little on FL, except for the EIA analysis~\cite{eia_florida_covid:20} that indicated the trend in FL was distinct from other studies.  




Assessing the impact of the pandemic on electricity consumption requires correcting for weather, so that the weather-independent part of demand can be extracted. It is not trivial to do so even for a single consumer~\cite{KissockAmbient:1998,BizEEDegreeDays:2020}.  Correcting for weather on aggregate demand in a large geographical area is far more challenging. The studies that corrected for weather had to make assumptions and approximations to handle this geographic variability~\cite{GuangchunCrosssDomainArXiV:2020,CicalaEarly:2020}. 
Our analysis of the three large regions (NY, CA, FL) is therefore based on the raw data to avoid the uncertainty introduced by the model used for weather correction. 
In the last section of the paper we perform weather correction of electricity demand from a small balancing authority in Florida for which weather can be defined clearly. In this particular area, weather corrected demand shows an increase due to the pandemic's mitigation efforts. 
This weather correction exercise also reveals the limit on the conclusions that can be made about the effect of the pandemic on electricity demand.

The rest of the paper is organized as follows. Section~\ref{sec:demand} analyzes electricity demand trends to identify impact of the pandemic on demand. Section~\ref{sec:pdfs} analyzes four variables (peak demand, demand ramp rate, forecast error and interchange) in order to assess if the pandemic is changing the stress on the grid. Section~\ref{sec:controlling-for-weather-method} presents a weather correction method and its application.




\section{Electric energy demand trends}
\label{sec:demand}
\begin{figure*}[htb]
\centering
\includegraphics[width=0.9\textwidth]{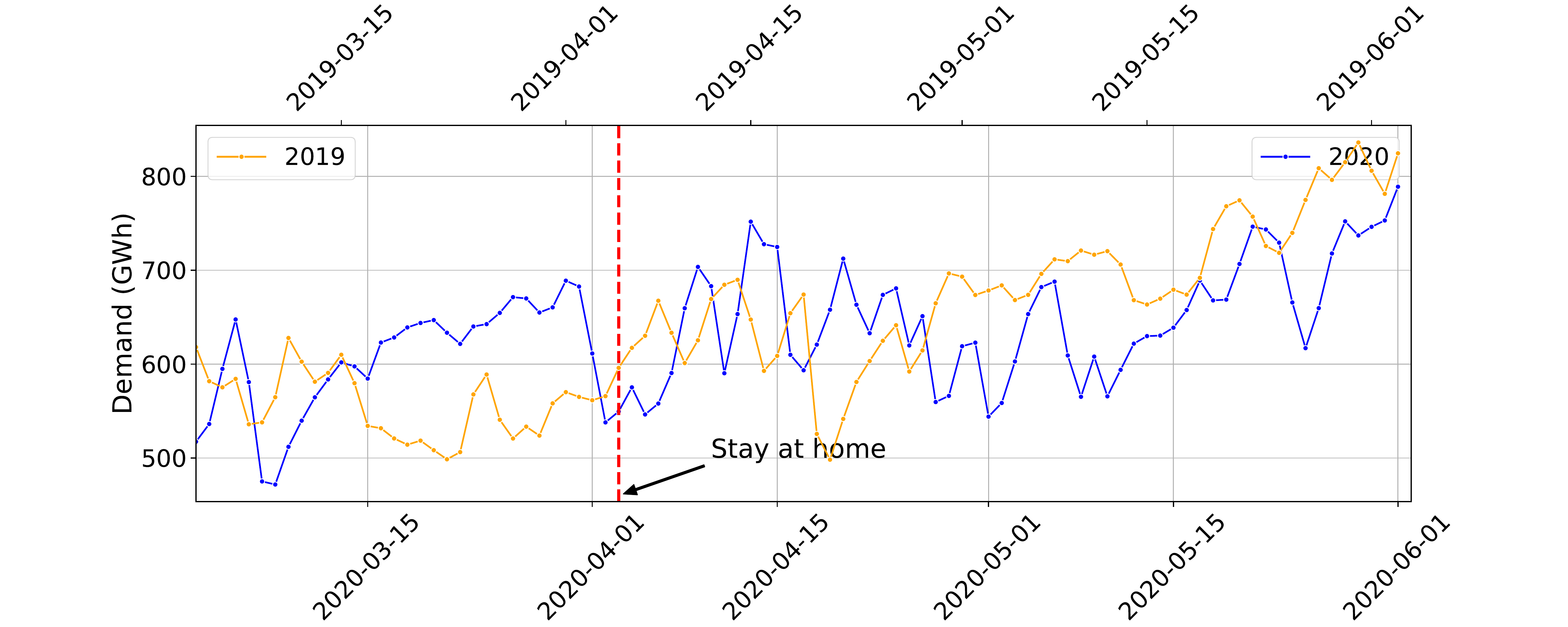} 
\caption{Daily electricity demand for Florida for March-May 2019 and 2020 (The first Monday of March for both years are vertically aligned.) }
\label{fig:TotalDailyDemandFL}
\end{figure*}

The daily electricity demand trend in Florida is shown in Figure~\ref{fig:TotalDailyDemandFL}. Apart from a temporary drop just before and after the statewide stay-at-home order, daily demand has an overall increasing trend across the two month period of March-April, 2020. The trend also does not seem to be correlated to that in the same period in 2019. The lack of a clear trend indicates that if the pandemic has had an impact on electricity demand, it will only become clear after the effect of weather is accounted for. 

\begin{figure*}[hbt]
\centering
\includegraphics[width=0.9\textwidth]{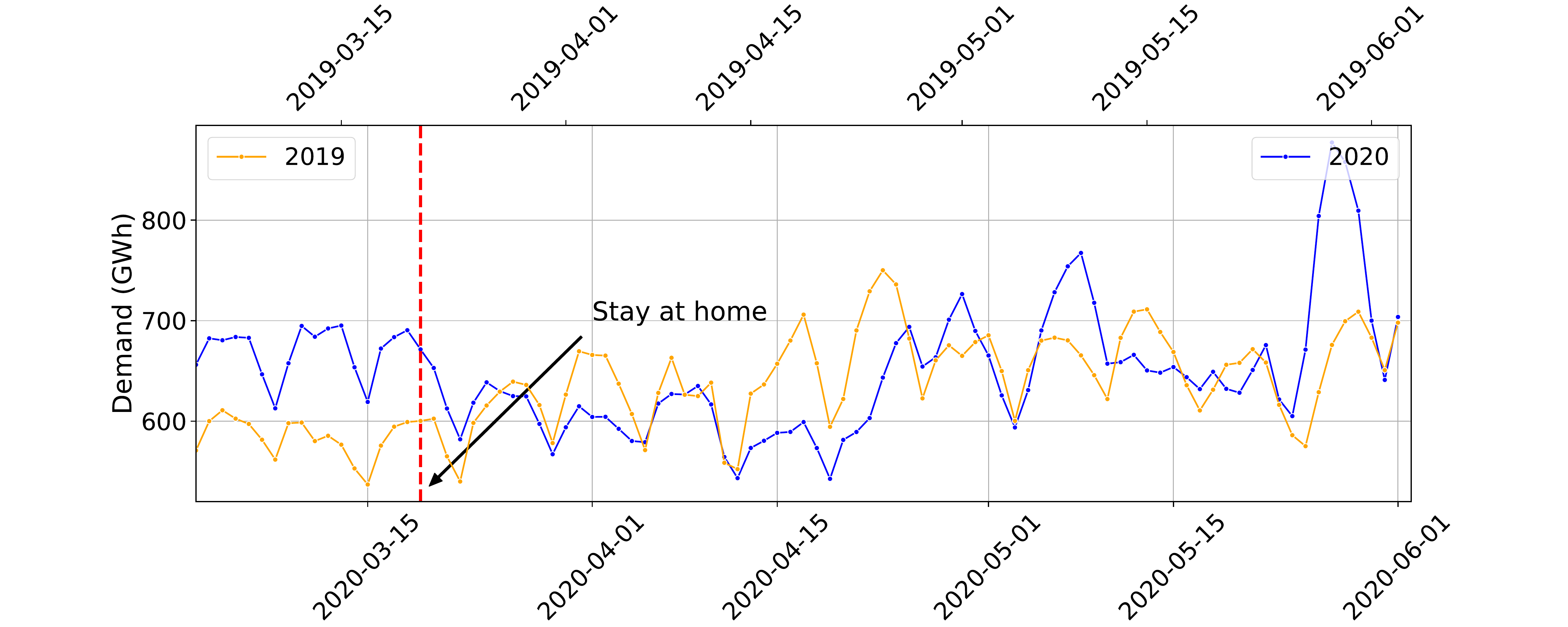} 
\caption{Daily Electricity Demand for California for March-May 2019 and 2020 (The first Monday of March for both years are vertically aligned).}
\label{fig:TotalDailyDemandCA}
\end{figure*}

Figure~\ref{fig:TotalDailyDemandCA} compares the daily electricity demand in California for March-April 2020 and 2019. Electricity demand appears to have decreased after the State of California issued the stay-at-home order. This is consistent with the observations in the report~\cite{EPRICOVID_March27:2020}. However, the demand trend in subsequent weeks, particularly the last week of April, is not consistent with a reduction. The 2020 demand not only appears to have reached to pre stay-at-home order levels, but also closed the gap with 2019 demand levels.

\begin{figure*}[htb]
\centering
\includegraphics[width=0.9\textwidth]{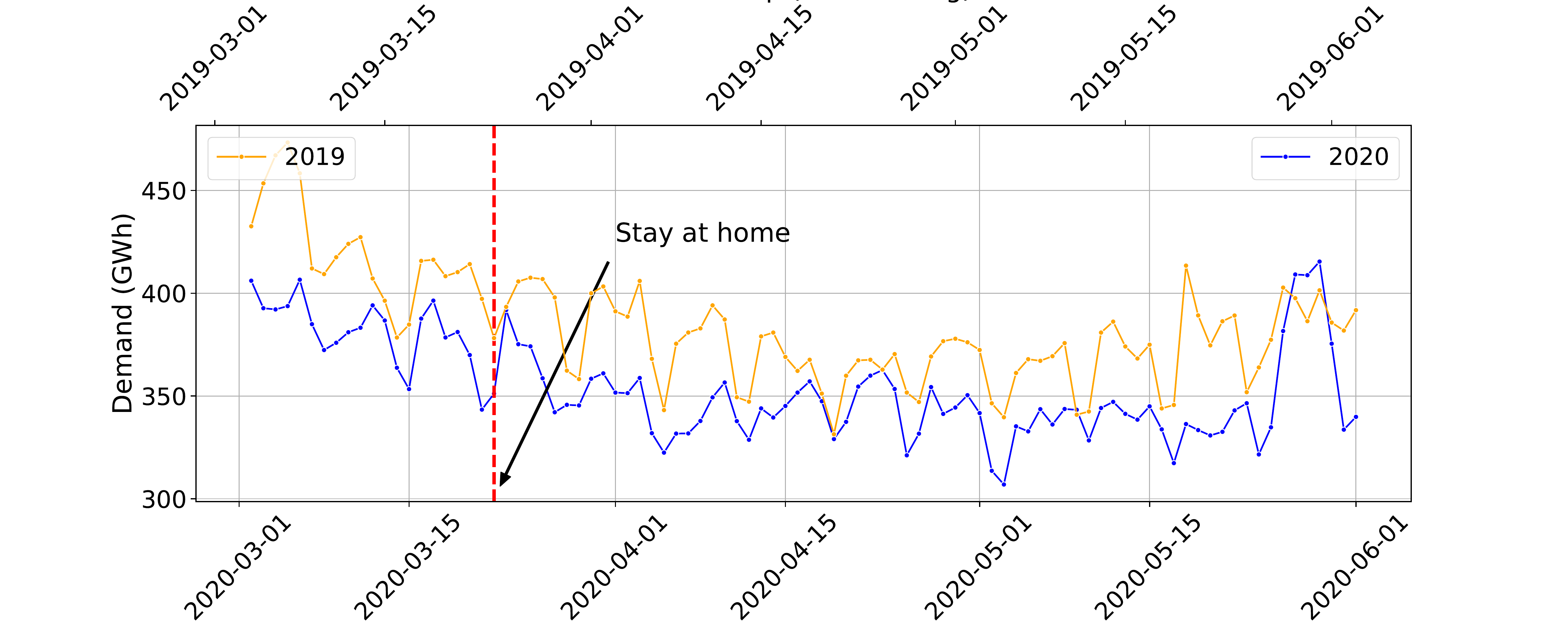} 
\caption{Daily electricity demand for New York for March-May 2019 and 2020 (The first Monday of March for both years are vertically aligned.)}
\label{fig:TotalDailyDemandNY}
\end{figure*}

The daily electricity demand in New York is shown in Figure~\ref{fig:TotalDailyDemandNY}. At first glance it appears to be consistent with the findings of the earlier studies~\cite{EPRICOVID_March27:2020,CicalaEarly:2020} (the latter was about Italy). However, further analyses - which we describe below - indicated that it will be inaccurate to infer from this data that the pandemic caused the demand reduction.

First, we compared the daily electricity demand for the state of New York from January to May for 2019 and 2020 (figure not shown). The data shows that the electricity demand for 2020 has been consistently lower than 2019, meaning the reduction could be due to factors other than the pandemic, such as weather.

Second, to further assess whether these trends are significant or not, we fitted simple linear regression models to data from both years. The models were daily electricity demand regressed over an array representing days passed from the first Monday of January for each month. Both models are significant and almost identical in parameter estimates. 

Thus, comparisons with weeks immediately before and after the stay-at-home order to identify impact of the pandemic should be done with care. No clear change is apparent after stay-at-home order in NY. Rather, the post-lockdown trend seems to be a continuation of the pre-lockdown trends. The overall conclusion about daily electricity demand is that the effect of the pandemic, if there is an effect at all, is not easily seen without  correcting for weather. 

\section{Analysis of variables that can reveal stress on the power grid}\label{sec:pdfs}
Reliable electricity supply requires a healthy power grid, which requires maintaining demand-supply balance at all times. Therefore, we analyze a number of additional variables which can impact demand-supply balance directly or indirectly, with a goal to detect changes due to the pandemic\footnote{Many of these variables, such as peak demand, should have the unit of power (MW or GW), not the unit of energy (MWh or GWh). However, since the only data we have are hourly energy data, we have defined all quantities in terms of hourly energy use, and thus they all have the unit of energy.}. Any significant change in these variables will indicate changes in the stress on the nation's power grid.

In the following, $d_k$ denotes the electricity demand (MWh) at hour (ending) $k$, where $k=1,\dots,24$ is an hour counter.
\begin{enumerate}
\item \emph{Daily peak and trough of the hourly demand}, defined as the maximum and minimum of the hourly electricity demand over a 24-hour period. A large value of the peak demand, particularly when it is significantly different from the base load, means additional peaker plants have to be brought on-line. Apart from increasing cost, reliance on peaker plants reduces reliability since grid operation becomes sensitive to potential failure or unavailability of these plants\footnote{Unavailability can occur due to various reasons, not just mechanical problems at the plant. For instance, if the grid operator does not expect the demand to be large it may not offer contracts ahead of time for the plants to be ready. }. 
\item \emph{Demand ramp rate,} defined as the difference between the hourly demand in an hour and the previous hour, i.e., $d_k - d_{k-1}$. Rapid and large changes in demand makes it harder for grid operators to balance demand and supply since generators are limited in their ability to increase or decrease generation. A large demand ramp rate is, thus, a measure of added stress on the grid.
\item \emph{Day ahead demand forecast error,} defined as the difference between the actual demand in an hour and the forecasted demand for that hour computed a day ahead. Grid operators compute forecasts for hurly demand 24 hours in advance (among many other types of forecasts). Many aspects of grid planning and operation, including unit commitment, will be inadequate in case of large forecast errors. The grid operator then has to procure more generating resources in real time. Also, a change in the forecast error can indicate a change in consumer behavior, all else  remaining constant.
\item  \emph{Net interchange,} defined as the electric energy exchanged between a region and its neighboring  regions every hour through the transmission network interconnecting them. It is positive if  the net energy exported is greater than that imported. A large interchange, and large variations in interchange over time, puts more stress on the transmission lines. Thus, changes in the trend in interchange is also a sign of potential change in grid's operational condition.
\end{enumerate}

\subsection{Peak (and trough) demand}
\begin{figure*}[hbt]
\centering
\includegraphics[width=0.9\textwidth,clip=true, trim=0in .5in 0in .5in]{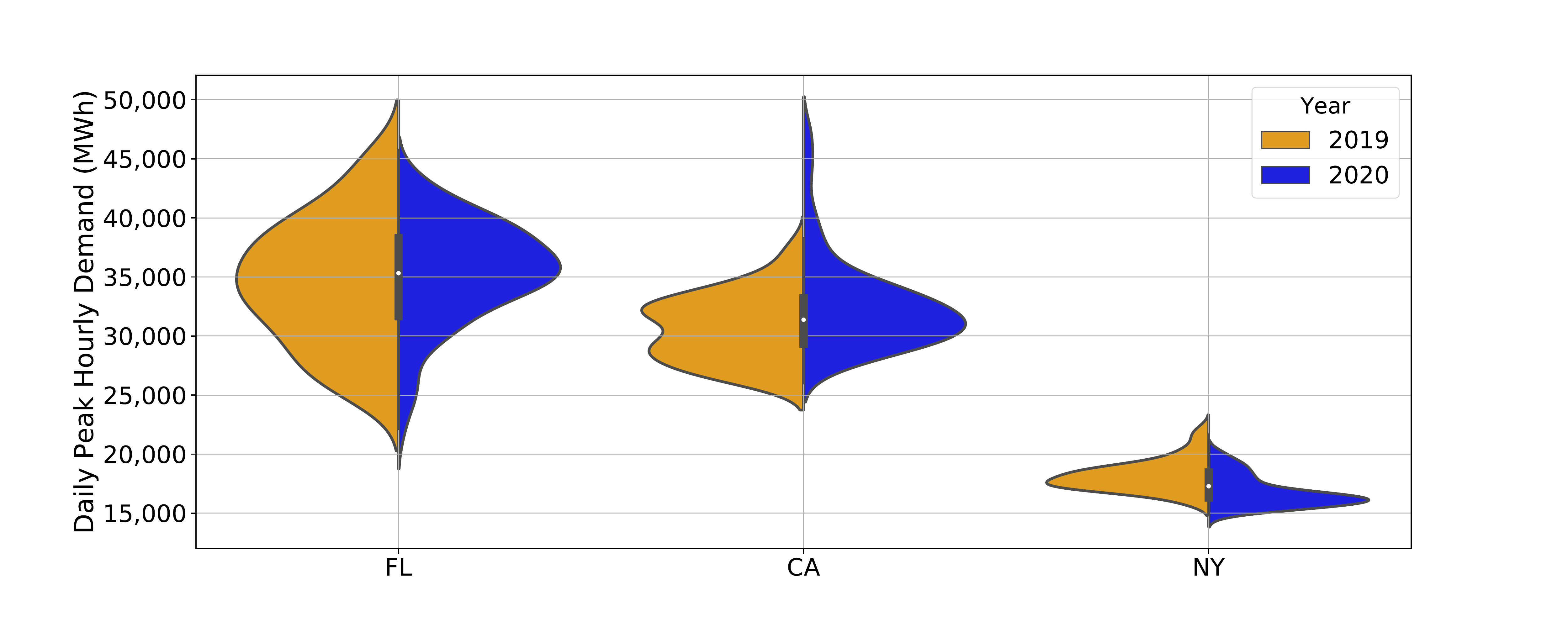} 
\caption{Distributions of daily peak demand (of hourly energy) in three regions in the US. Data from Jan - May, during 2019 and 2020, are used to estimate the pdfs.}
\label{fig:ViolinBA}
\end{figure*}

Figure~\ref{fig:ViolinBA} shows the so-called violin plots, i.e.,  empirical probability density functions (pdfs) estimates of the daily peak of the hourly demand (MWh per hour)  for three states: Florida, California and New York. Data from Jan-May of a year is used to estimate the pdf, using the kernel density estimator with a Gaussian kernel. 

We see from the plot that both in California and New York have lower extreme values of the peak demand in first quarter of 2020 when compared to 2019, but the trend is the opposite in Florida. If peak demand was the sole determinant of the stress on the grid, we can conclude that the stress on grid operation has reduced in California and New York, but the opposite has happened in Florida.

\begin{figure*}[ht]
\centering
\includegraphics[width=0.9\textwidth,clip=true, trim=0in 1.5in 0in 1.5in]{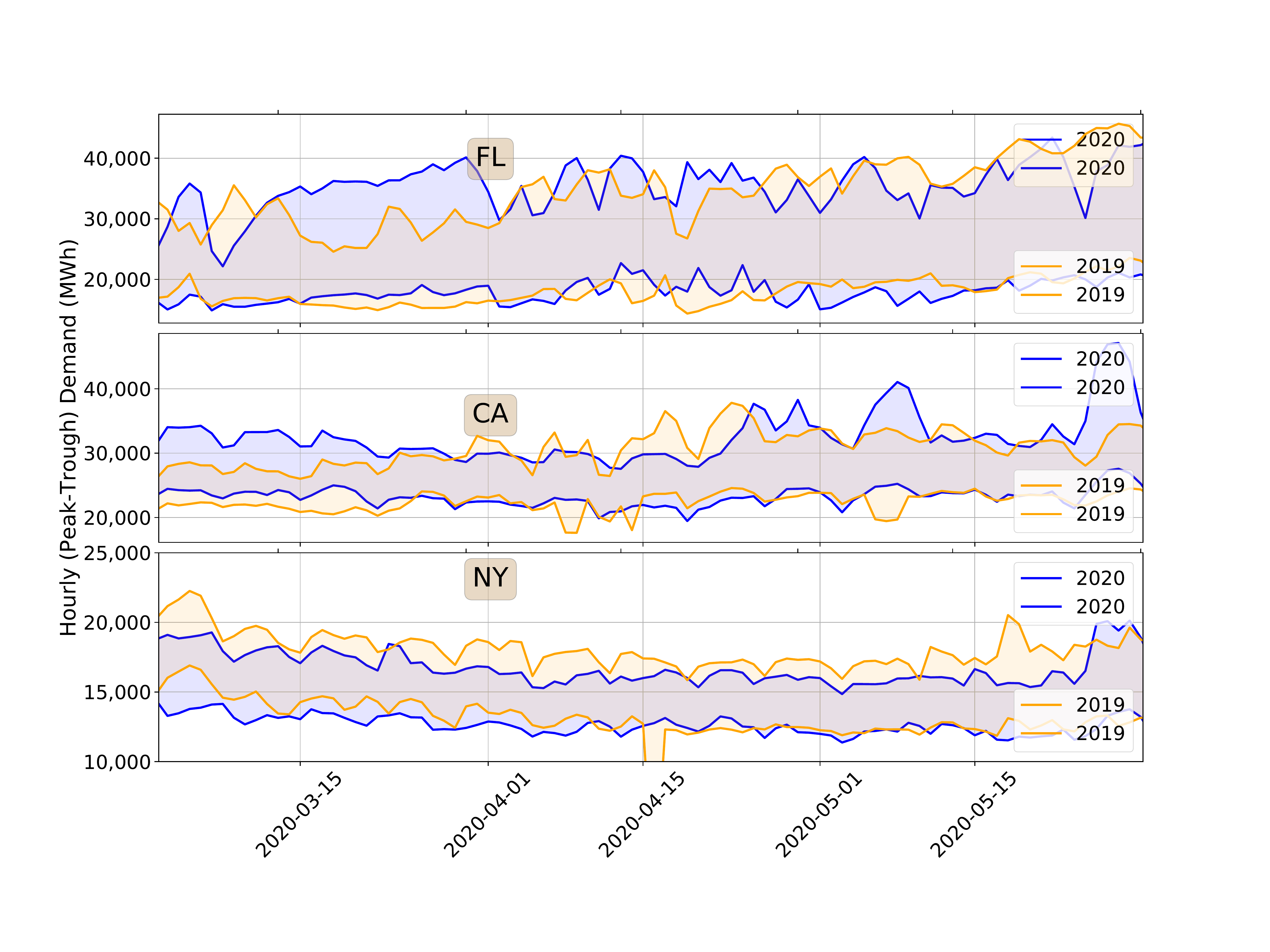}  
\caption{Daily peak-trough of hourly demand for each day, March-May 2019 and 2020. (The first Monday of March 2019 is aligned with the first Monday of March 2020.)}
\label{fig:DemandVariability}
\end{figure*}

To check if there is a change in the peak demand because of the pandemic, or is the change routinely occurring throughout January-May 2020, we also look at the data in time domain. Figure~\ref{fig:DemandVariability} shows the daily peak and trough of the hourly electricity demand for Florida, California and New York. In California and Florida, both the peak and trough were greater in March 2020 than in March 2019. Following stay-at-home orders, both peak and trough has decreased making their difference comparable to those observed in 2019. But then it changed back to pre-pandemic levels in late April.

The trend in New York is different from those in California and Florida. The peak and trough of hourly demand decreased consistently through March and April, in both 2018 and 2020. Also, it did not show the ``springing back'' effect seen in Florida and California until the end of May (2020). We also extended the analysis period to January, and a similar trend to daily total demand was observed here. Both peak and trough demand for 2019 and 2020 steadily declined from early January to the late May.

\subsection{Demand ramp rate}
\begin{figure*}[htb]
\centering
\includegraphics[width=0.325\textwidth]{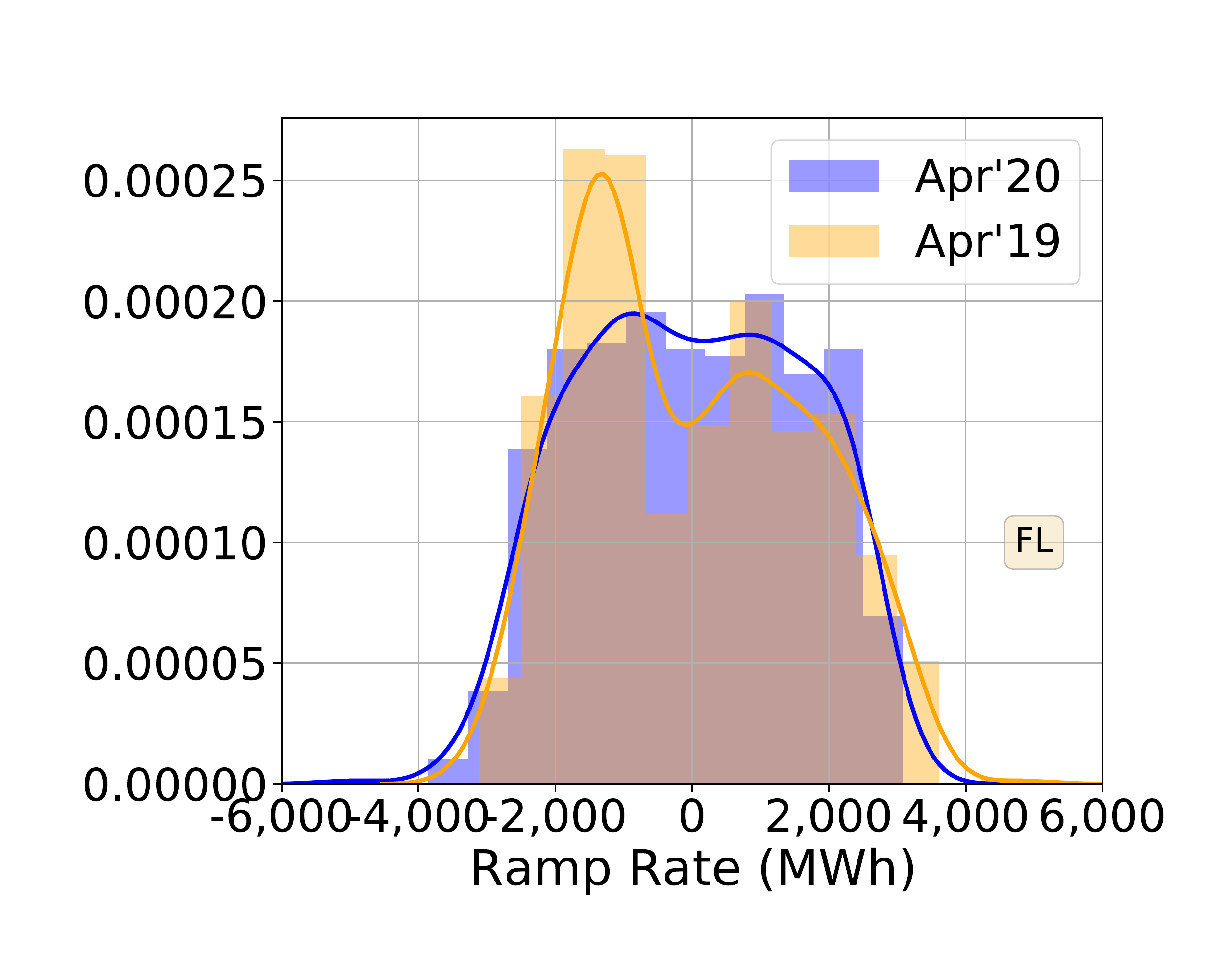} 
\includegraphics[width=0.325\textwidth]{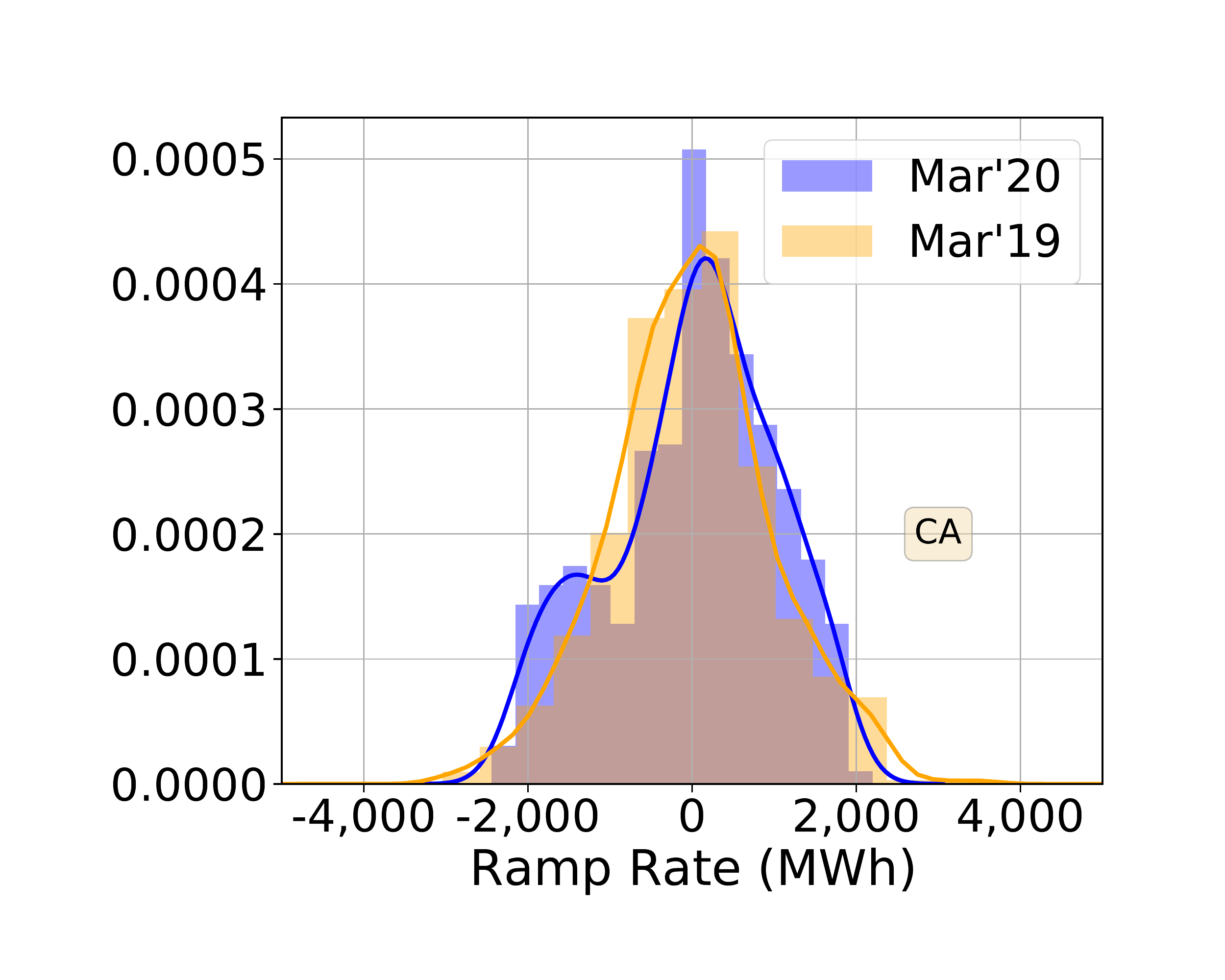} 
\includegraphics[width=0.325\textwidth]{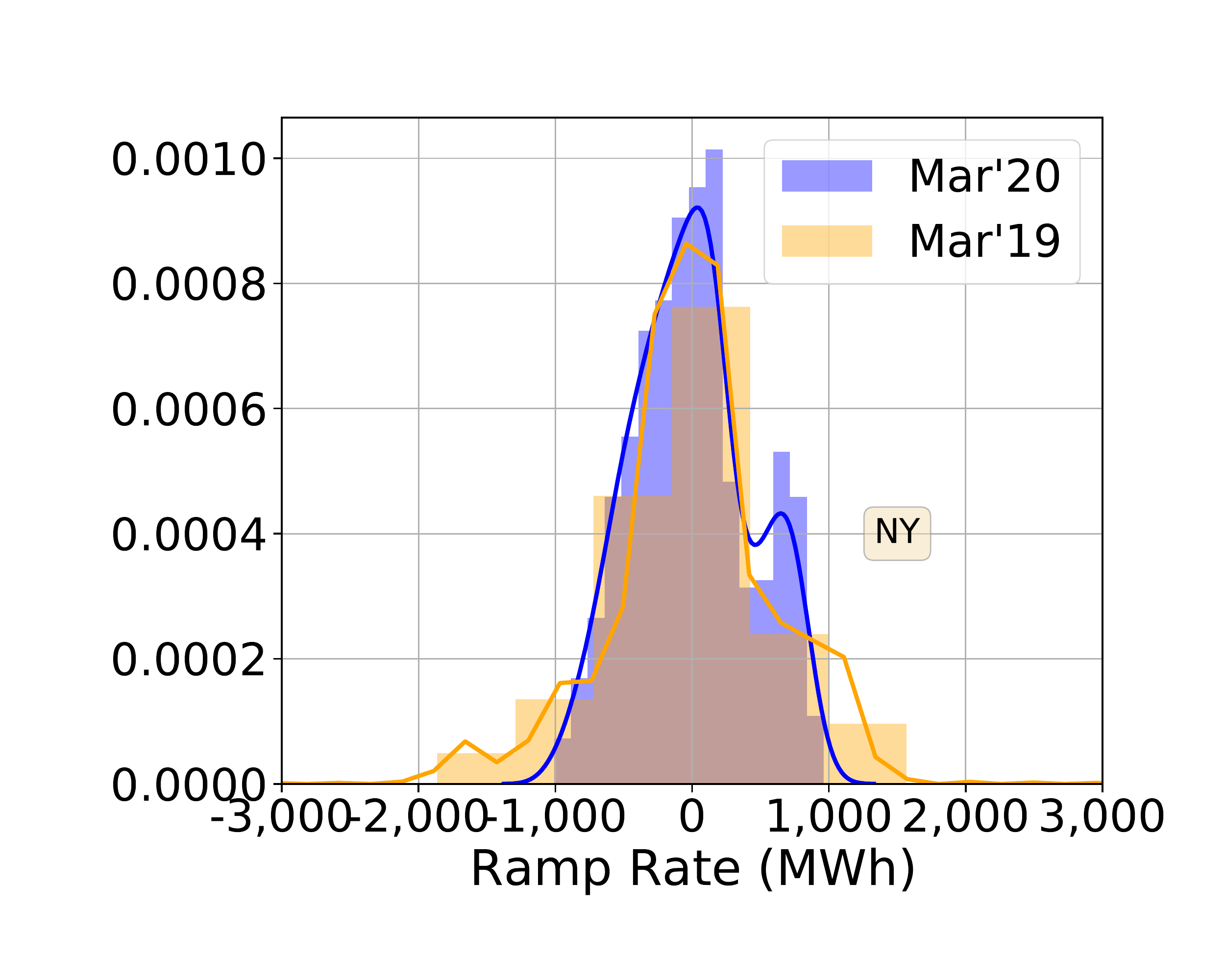} 
\caption{Probability densities of hourly demand ramp rate, with 4 weeks of data after the stay-home order in 2020, and the corresponding period in 2019.}
\label{fig:ramprate-onpandemic}
\end{figure*}

Figure~\ref{fig:ramprate-onpandemic} compares the empirical probability density functions (pdfs) of the hourly demand ramp rate for the three states in 2020 (after the pandemic) and 2019 (corresponding period). The pdfs are estimated using data for a four week period, starting on the statewide stay-at-home order for the corresponding state, and for the same period in 2019. In Florida, although the pdf has changed after the stay-at-home order compared to the same period in 2019, the frequency of larger values did not  change significantly. There is no clear change in California. In New York, larger ramp rates are observed less frequently after the pandemic in 2020 than in 2019, meaning hourly demand is changing more slowly. From the point of view of grid operation, that is a positive change.

\begin{figure*}[htb]
  \centering
  \subcaptionbox{Ramp rate pdfs, pre-pandemic 2020 (January) and corresponding period in 2019. \label{fig:ramprate-prepandemic}}{%
\includegraphics[width=0.325\textwidth]{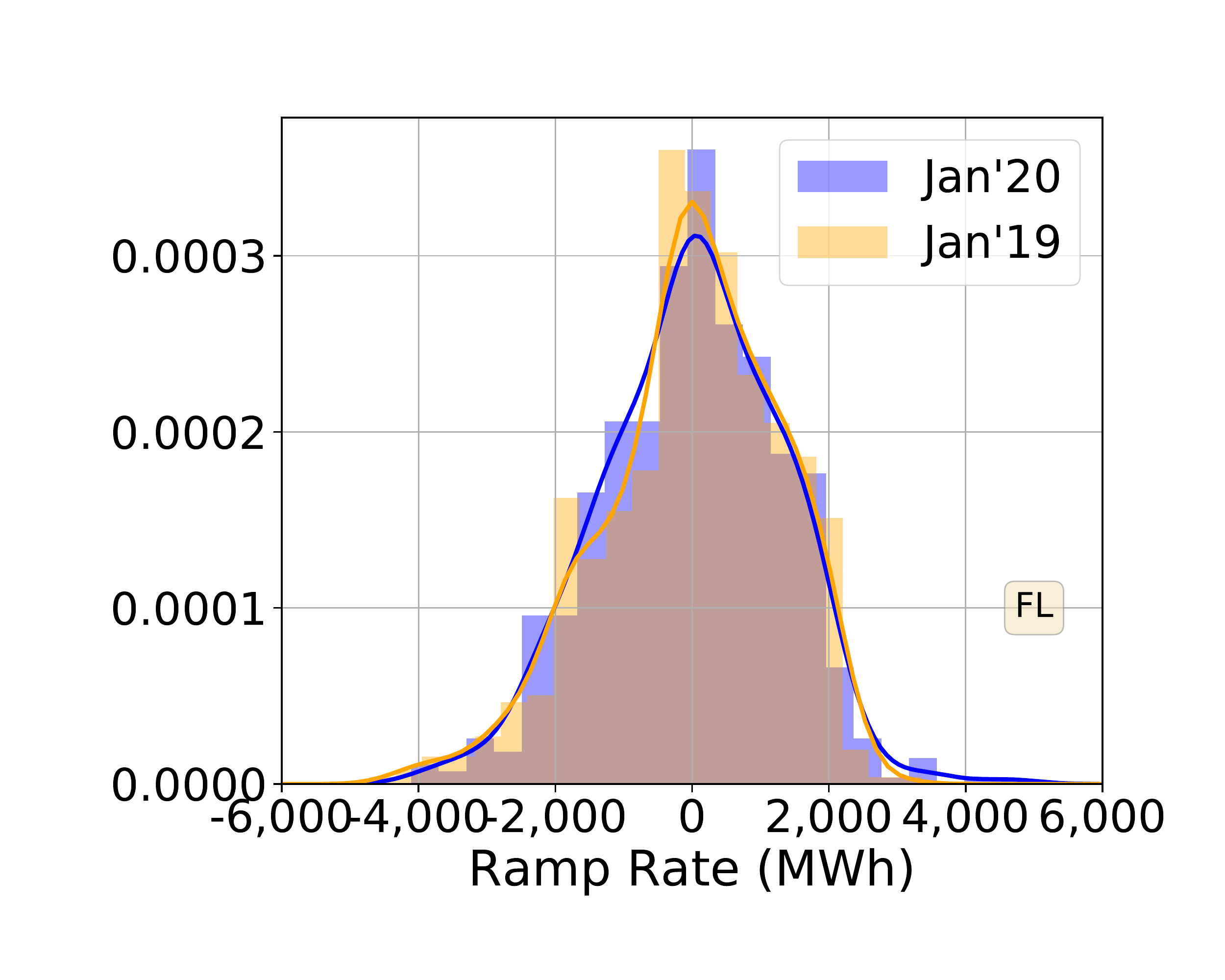} 
\includegraphics[width=0.325\textwidth]{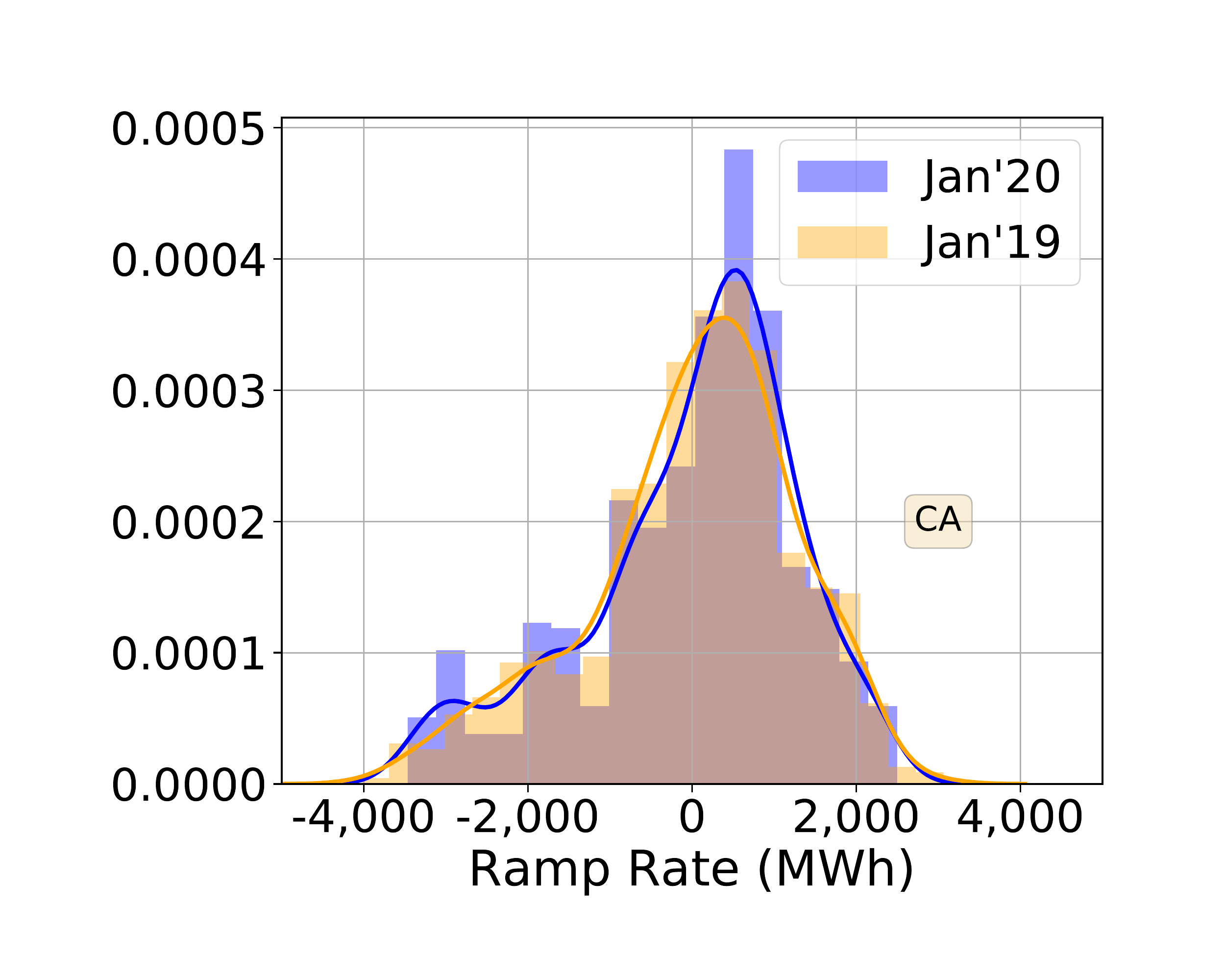} 
\includegraphics[width=0.325\textwidth]{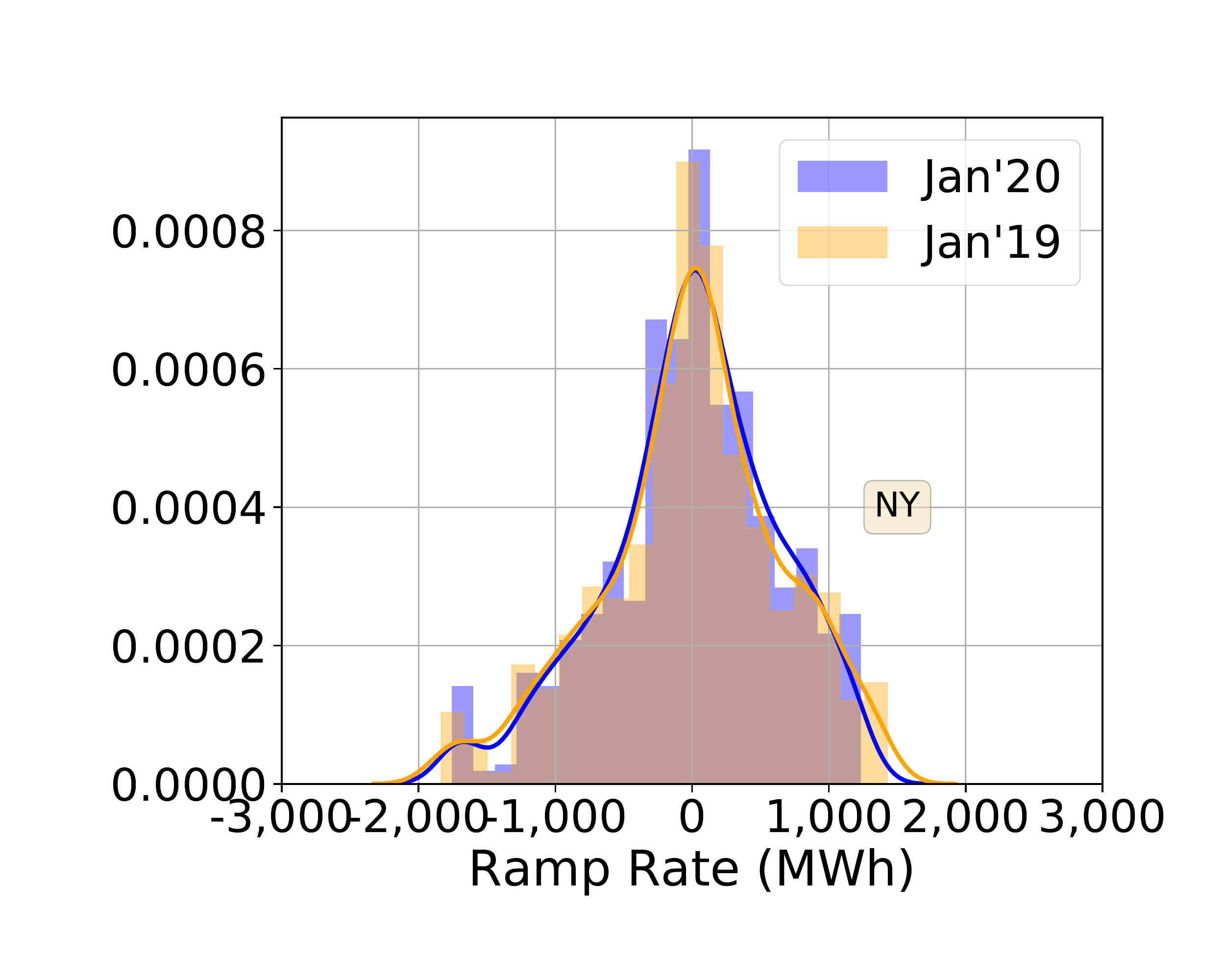} 
}
\subcaptionbox{Ramp rate distribution, post-pandemic 2020 (May) with corresponding period in 2019\label{fig:ramprate-May2020}}{%
    \includegraphics[width=0.325\textwidth]{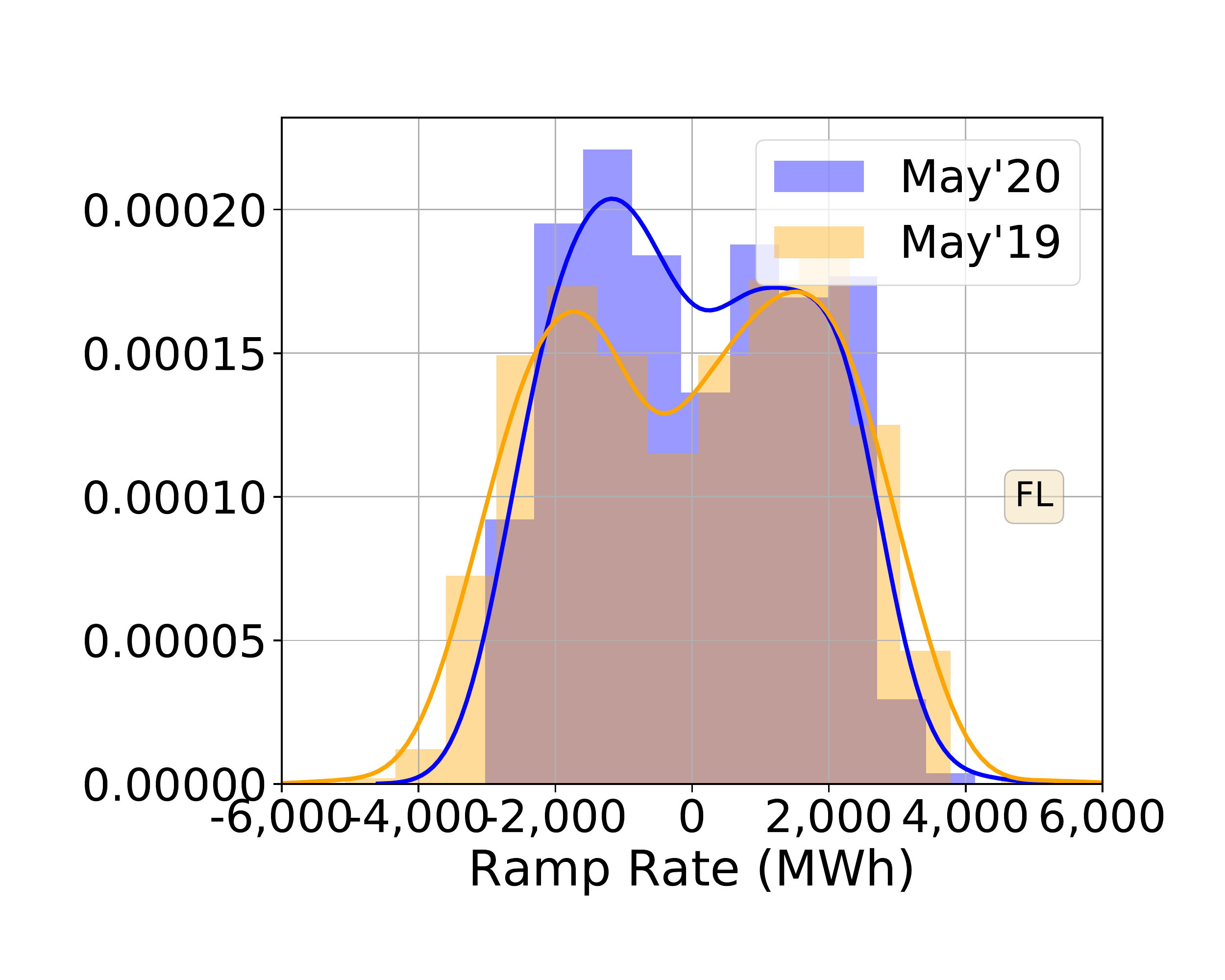} 
    \includegraphics[width=0.325\textwidth]{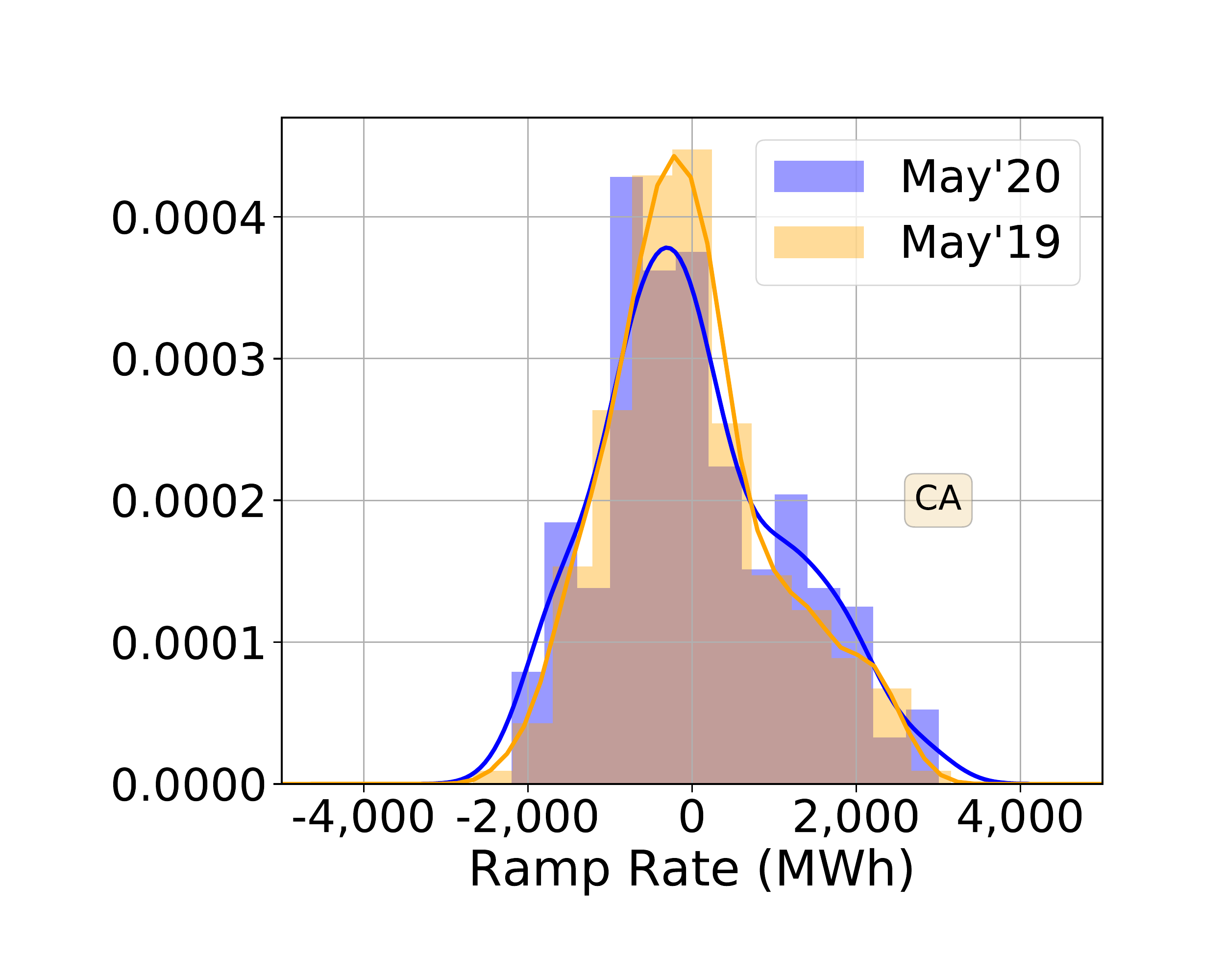} 
    \includegraphics[width=0.325\textwidth]{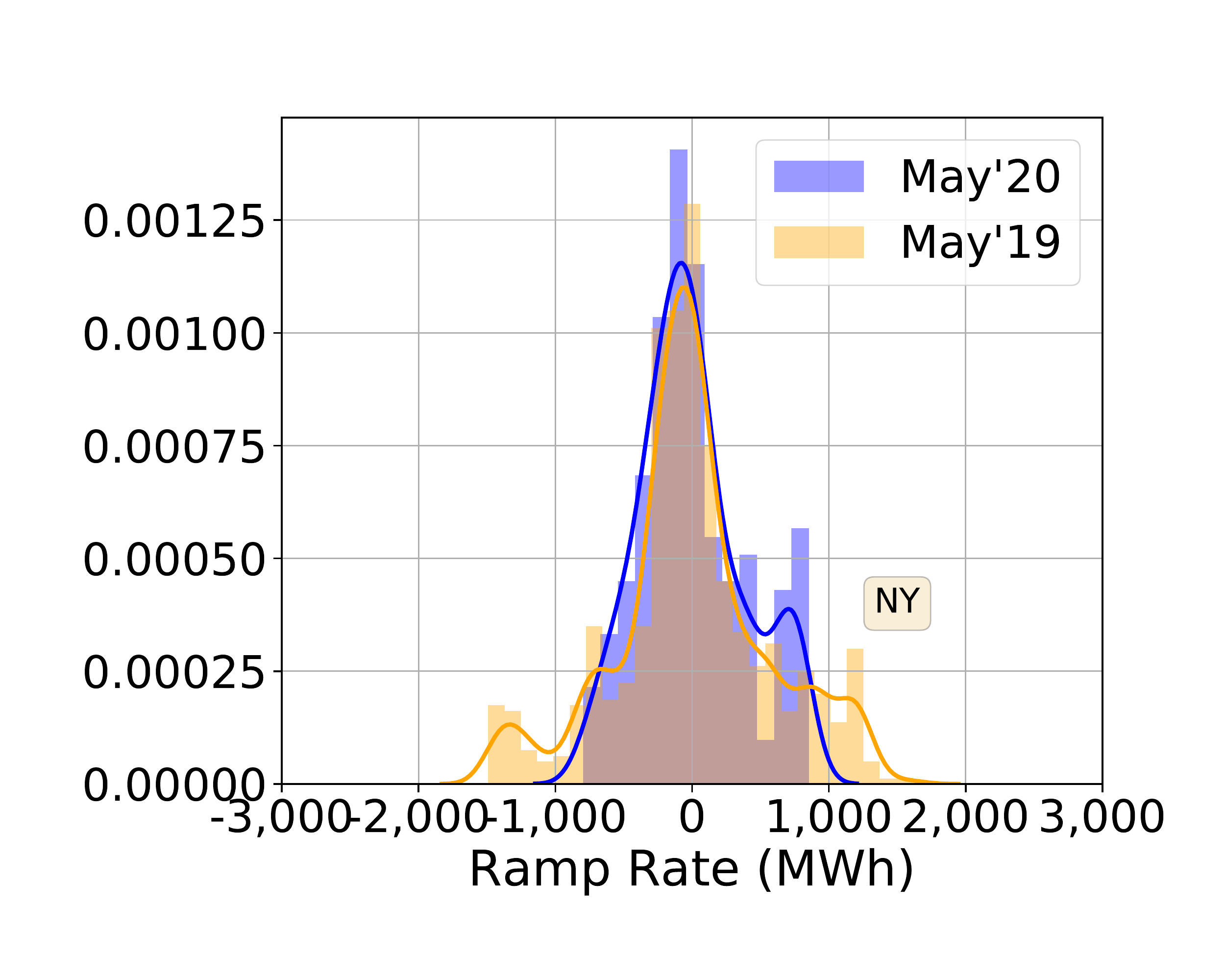} 
}
  \caption{Probability densities of hourly demand ramp rate, during months away from the pandemic (before and after).  }\label{animals}
\end{figure*}

 These trends have appeared after the pandemic, and are not part of a long term change in statistics of ramp rate. Figure~\ref{fig:ramprate-prepandemic}, which compares the pdfs in January 2020 with those in January 2019, support this observation. Ramp rate statistics during January 2020 are quite similar to those in 2019. Another observation is that the trends that appeared right after the pandemic have stayed that way and have not reverted back; see May 2020 pdfs in Figure~\ref{fig:ramprate-May2020}.
 
\subsection{Demand forecast error}
\begin{figure*}[ht]
\centering
\includegraphics[width=0.9\textwidth,clip=true, trim=0in 1.5in 0in 1.5in]{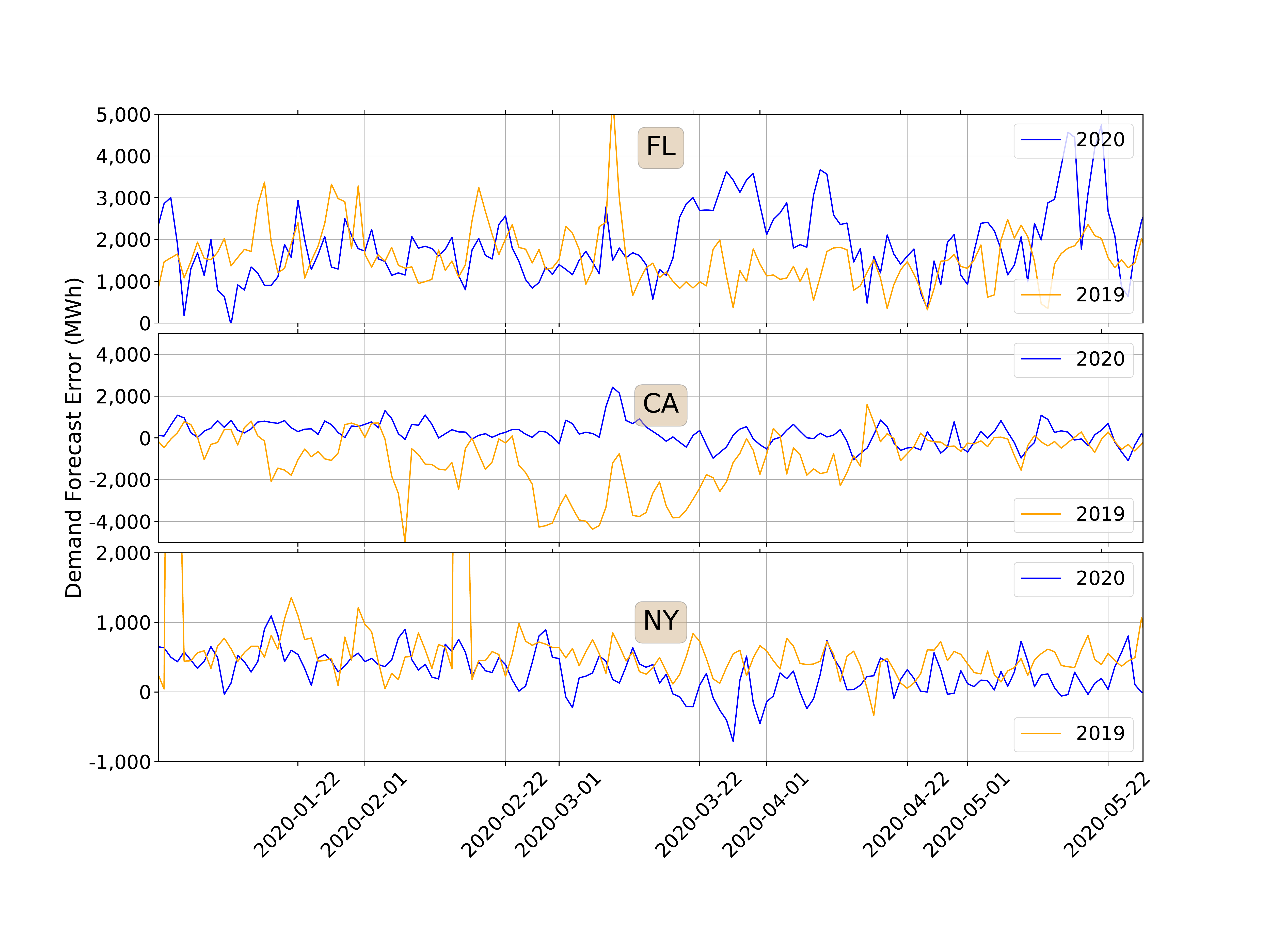}   
\caption{Daily average of the hourly demand forecast error,  2020 vs. 2019}
\label{fig:demandForeCastError-timedomain}
\end{figure*}

Recall that the hourly demand forecast error is $d_k - \hat{d}_k$, where $\hat{d}_k$ is the forecast of hourly demand $d_k$ computed 24 hours earlier. The time series of hourly forecast error is highly non-stationary. So we start with the daily average of the hourly data. Figure~\ref{fig:demandForeCastError-timedomain} shows the daily mean of the hourly forecast error, for 2020 and 2019 trends, from January to April. 

Figure~\ref{fig:demandForeCastError-timedomain} shows that in California, the mean forecast error is considerably smaller in 2020 than in 2019. It appears that 2019 was an outlier year: the 2018 forecast errors were considerably smaller than those in 2019 as well.  This makes it impossible to compare the statistics after the pandemic with those from the corresponding period in 2019 in any meaningful manner. Therefore, we ignore California in the rest of the discussion on forecast error, although we show the plots for the sake of completeness.

\begin{figure*}[ht]
\centering
\includegraphics[width=0.325\textwidth]{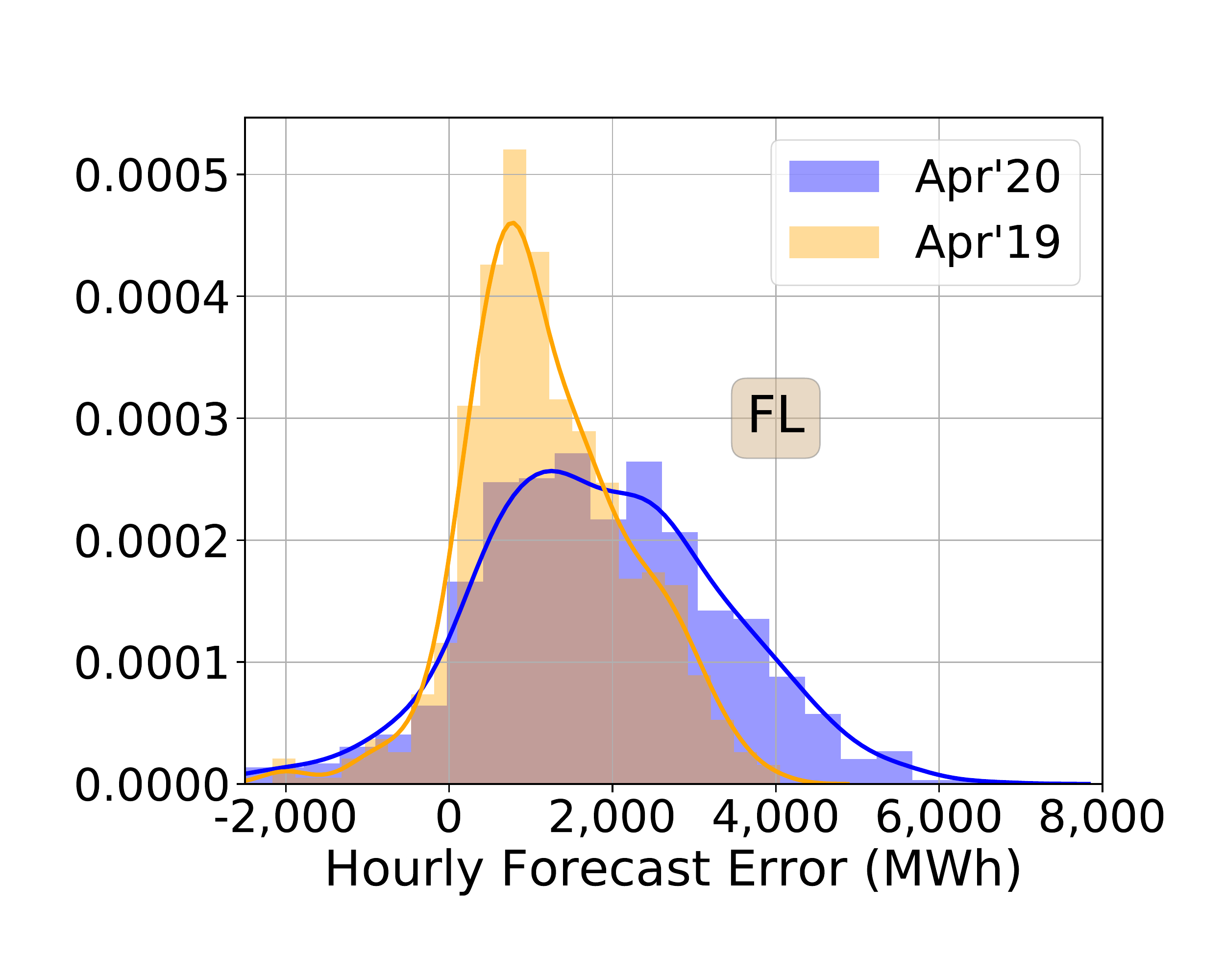}
\includegraphics[width=0.325\textwidth]{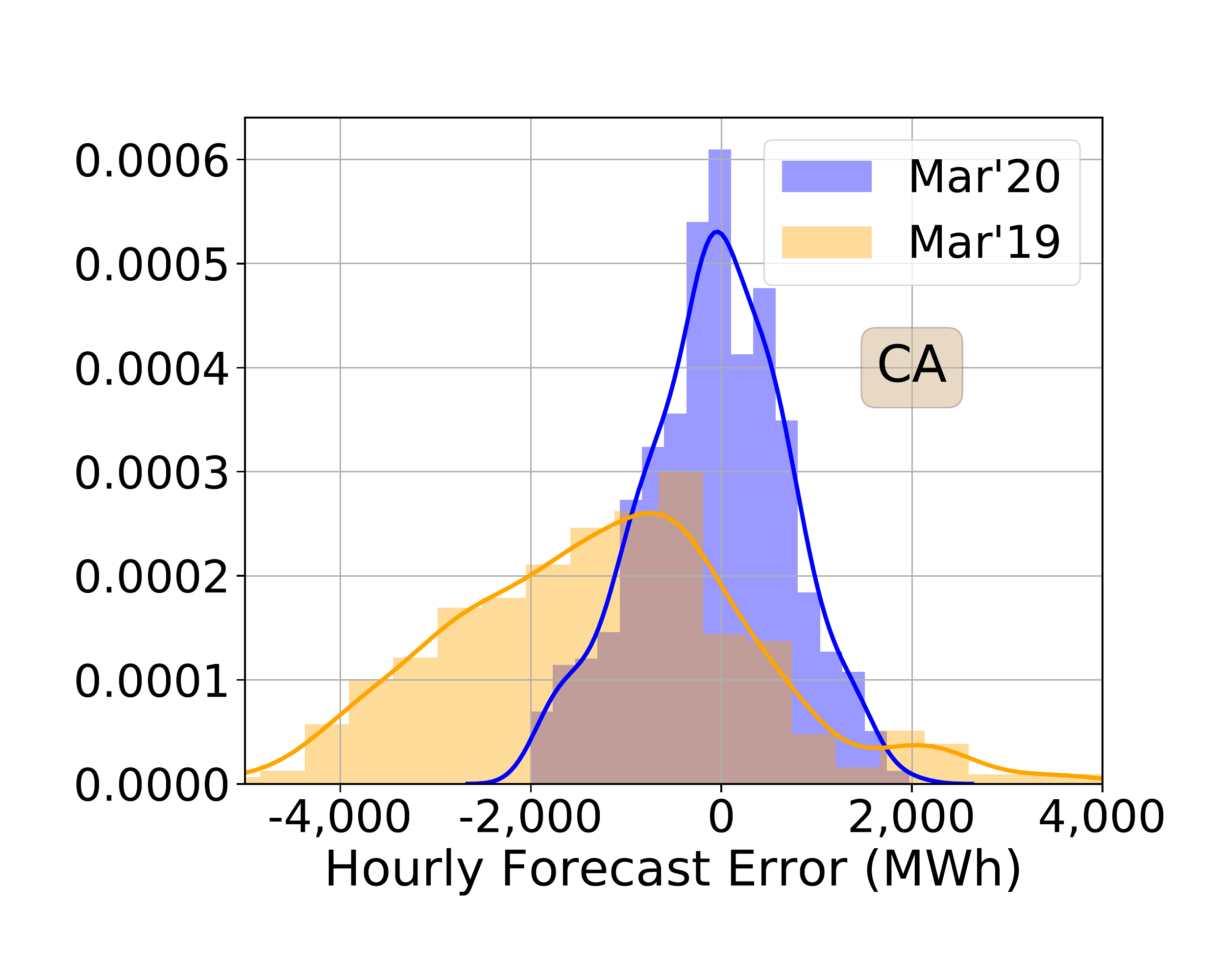}
\includegraphics[width=0.325\textwidth]{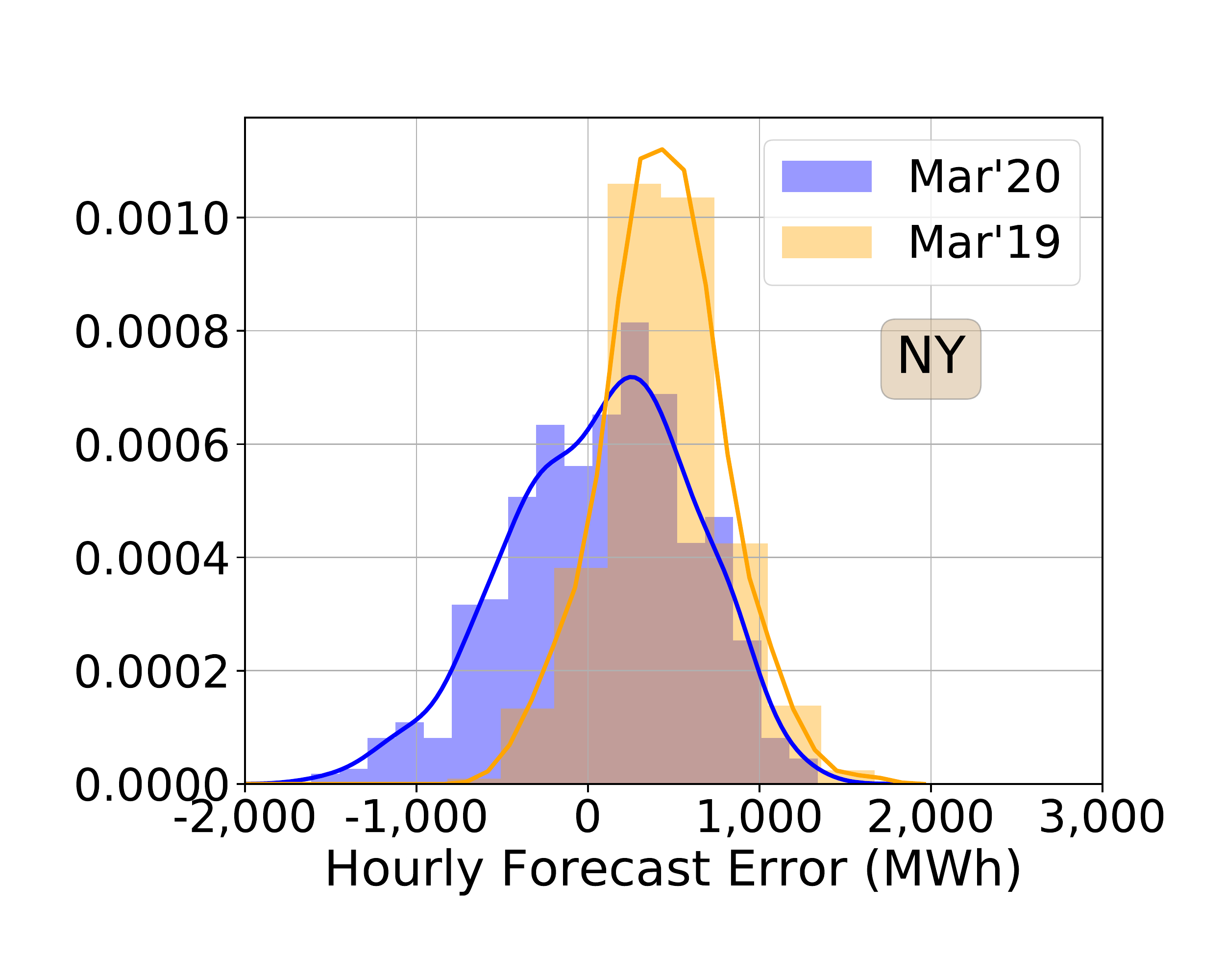} 
\caption{Hourly demand forecast error probability densities, with data from a four week period starting on the day stay-at-home order was issued. The 2019 data was for the corresponding period.}
\label{fig:demandForecastErrorHist-2020vs2019-onSAH}
\end{figure*}

For Florida and California, the mean forecast error does not show a clear trend that is correlated with the pandemic. 
 We, therefore, examine the empirical pdfs to get a more detailed picture of the demand forecast errors. Figure~\ref{fig:demandForecastErrorHist-2020vs2019-onSAH} compares the empirical pdfs of the hourly forecast error during a four-week period after the stay-at-home order in 2020 with those from the same period in 2019. There is a discernible difference between the 2020 and 2019 statistics---forecast errors have increased for both Florida and New York after the pandemic.

\begin{figure*}[ht]
  \centering
    \subcaptionbox{Hourly demand forecast error pdfs, pre-pandemic 2020 (January) and corresponding period in 2019. \label{fig:demandforecastHist-Jan2020-March2020}}{%
\includegraphics[width=0.325\textwidth]{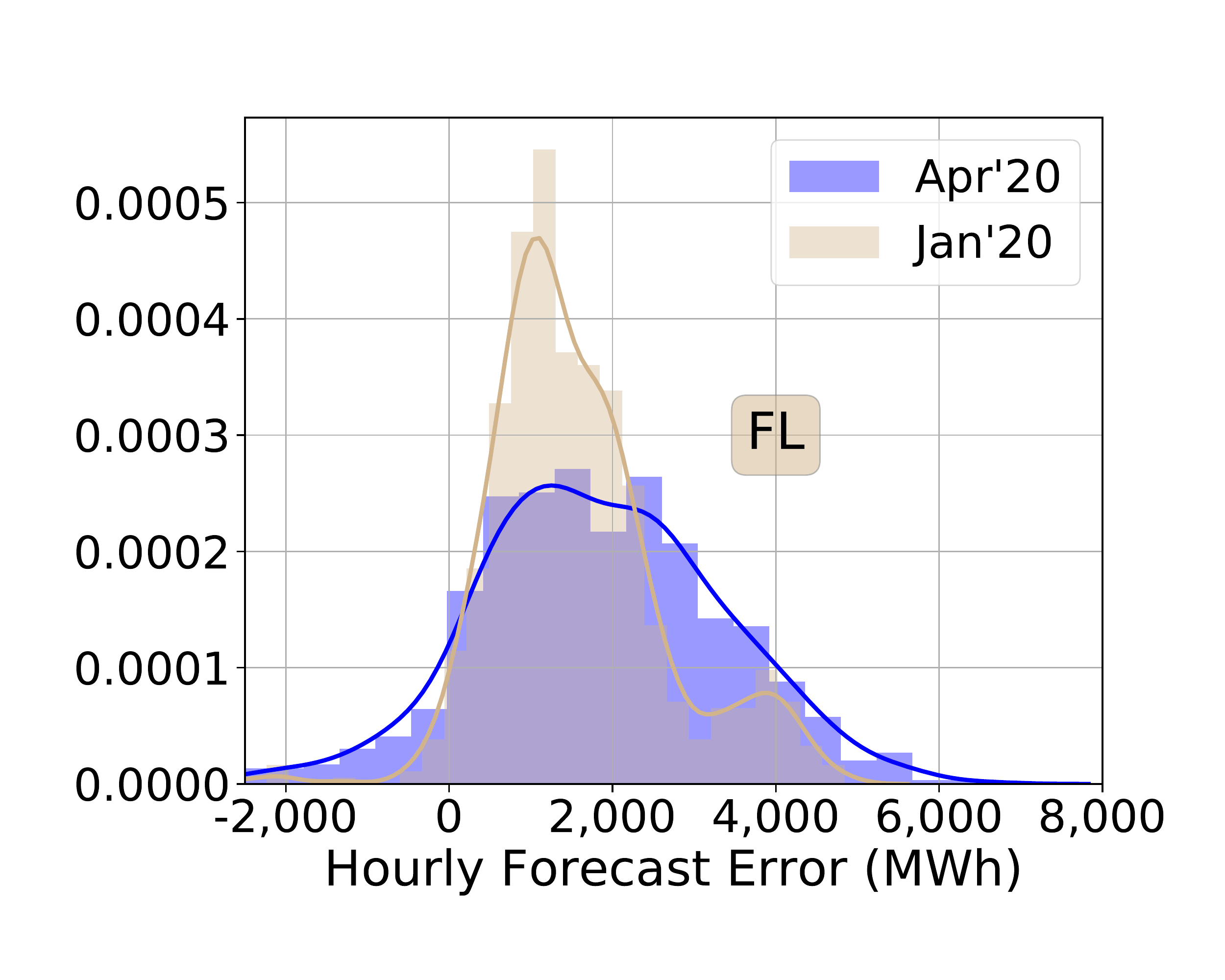}
\includegraphics[width=0.325\textwidth]{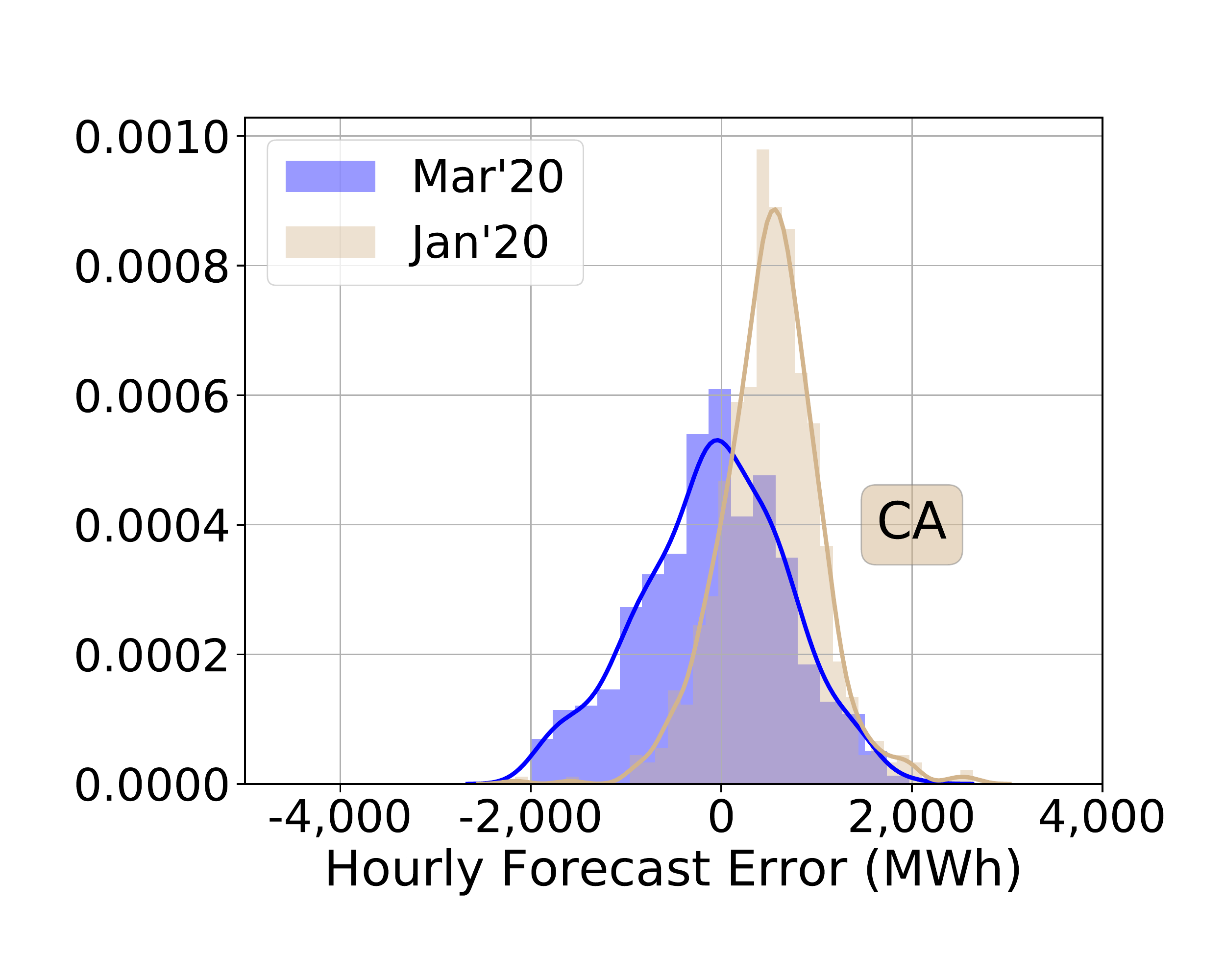} 
\includegraphics[width=0.325\textwidth]{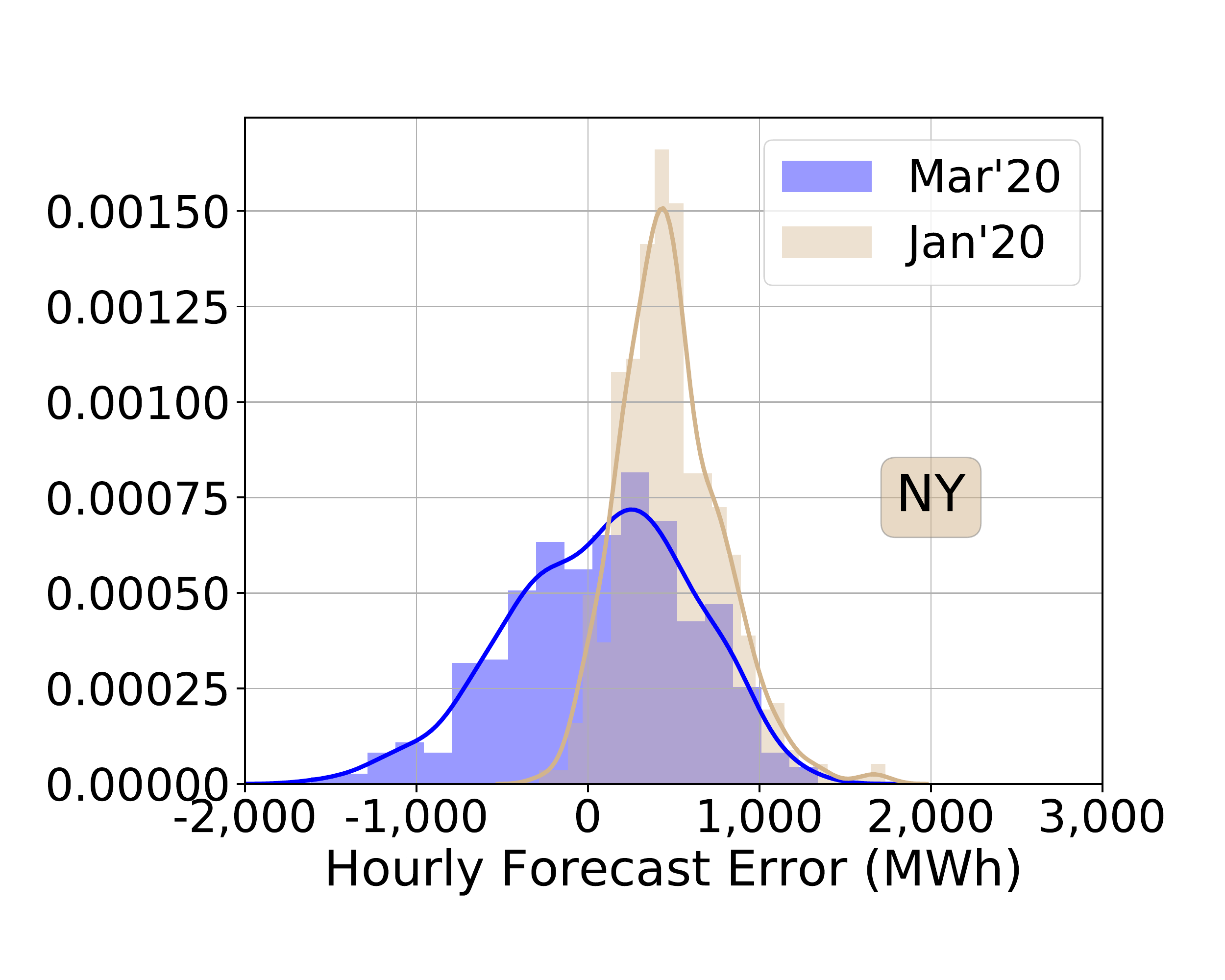}
}
  \subcaptionbox{Hourly demand forecast error pdfs, post-pandemic 2020 (May) and corresponding period in 2019. \label{fig:demandforecastHist-Jan2020-May2020}}{%
    \includegraphics[width=0.325\textwidth]{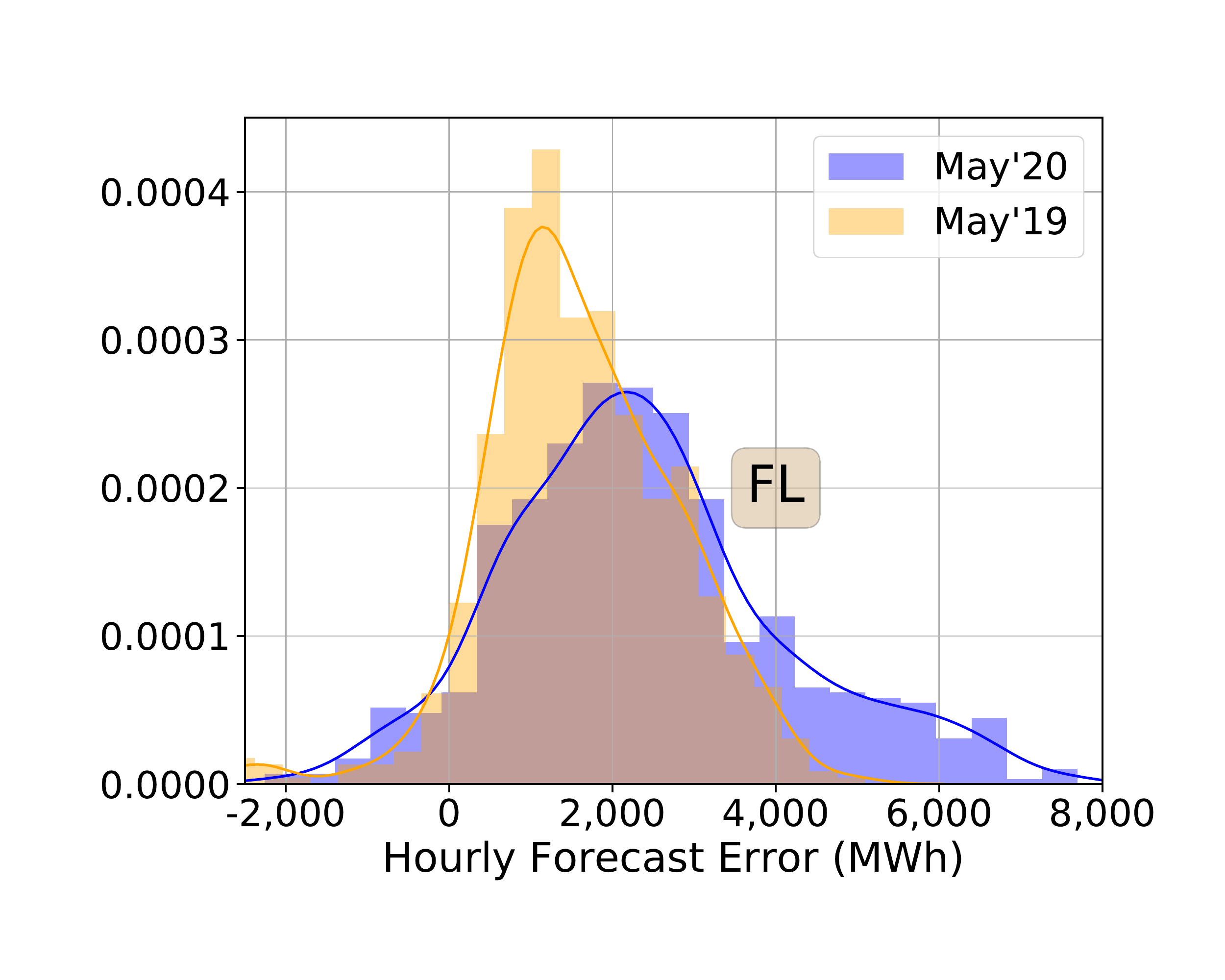}
\includegraphics[width=0.325\textwidth]{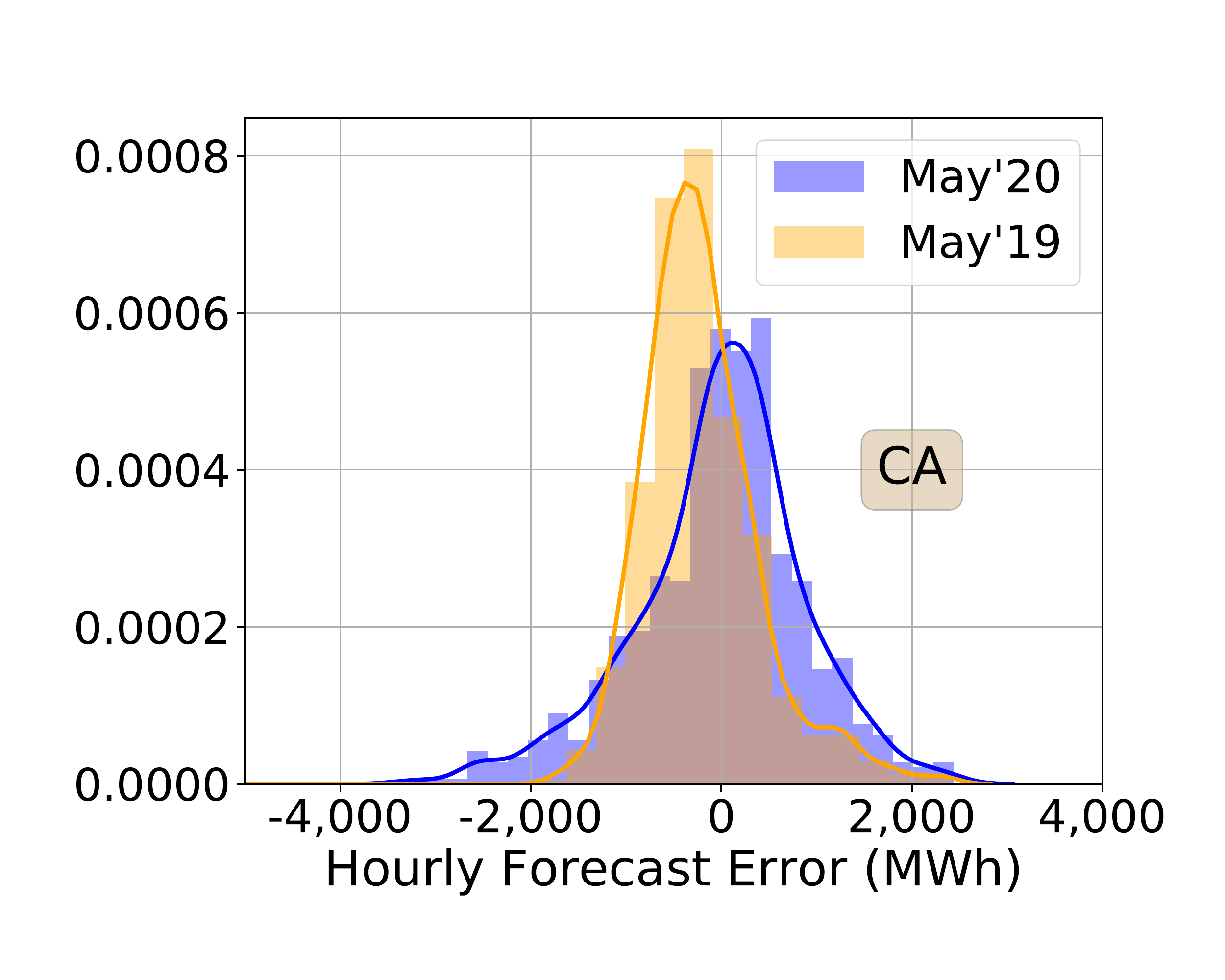} 
\includegraphics[width=0.325\textwidth]{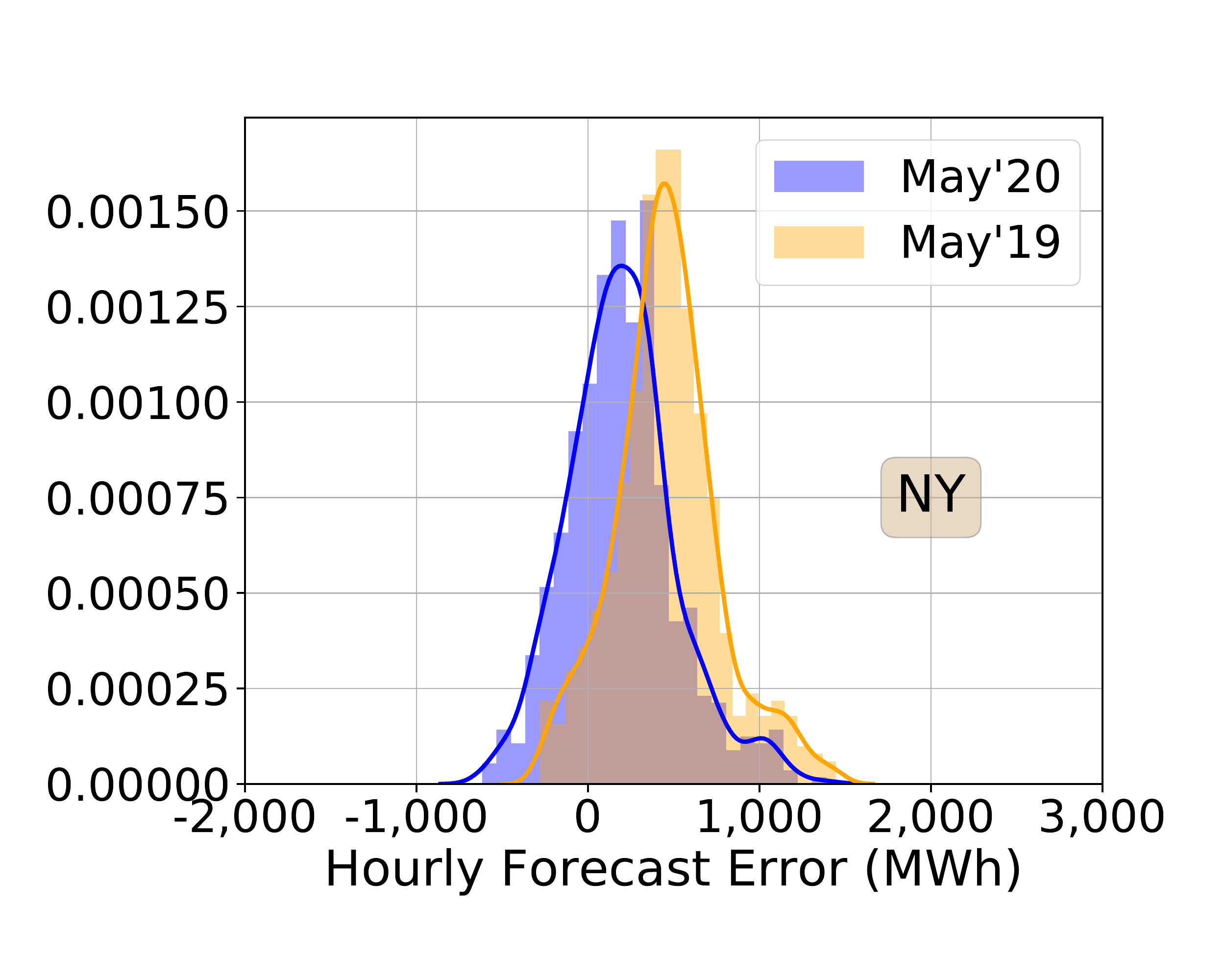}
}
\caption{Empirical probability densities of the hourly demand forecast errors. Each pdf is computed with data from a four week period, except for those for May 2020.}
\label{fig:forecastErrorHist-2}
\end{figure*}

To address the question about whether these changes are really due to the pandemic or simply a part of a long change between 2019 and 2020 already underway, we compare the pdfs within 2020, before and after the pandemic. It appears that the pandemic is the likely cause, as the forecast errors in January 2020, before the pandemic, were significantly smaller than in March/April (Figure~\ref{fig:demandforecastHist-Jan2020-March2020}). In addition, the forecast errors in NY show a clear ``springing back'' effect, but not in Florida (Figure~\ref{fig:demandforecastHist-Jan2020-May2020}).

Overall, the day ahead demand forecasts have been negatively affected by the pandemic in both Florida and New York. In New York, they have recovered to pre-pandemic levels by May but not in Florida.

\subsection{Total interchange}
\begin{figure*}[ht]
\centering
\includegraphics[width=0.9\textwidth,clip=true, trim=0in 1in 0in 1.5in]{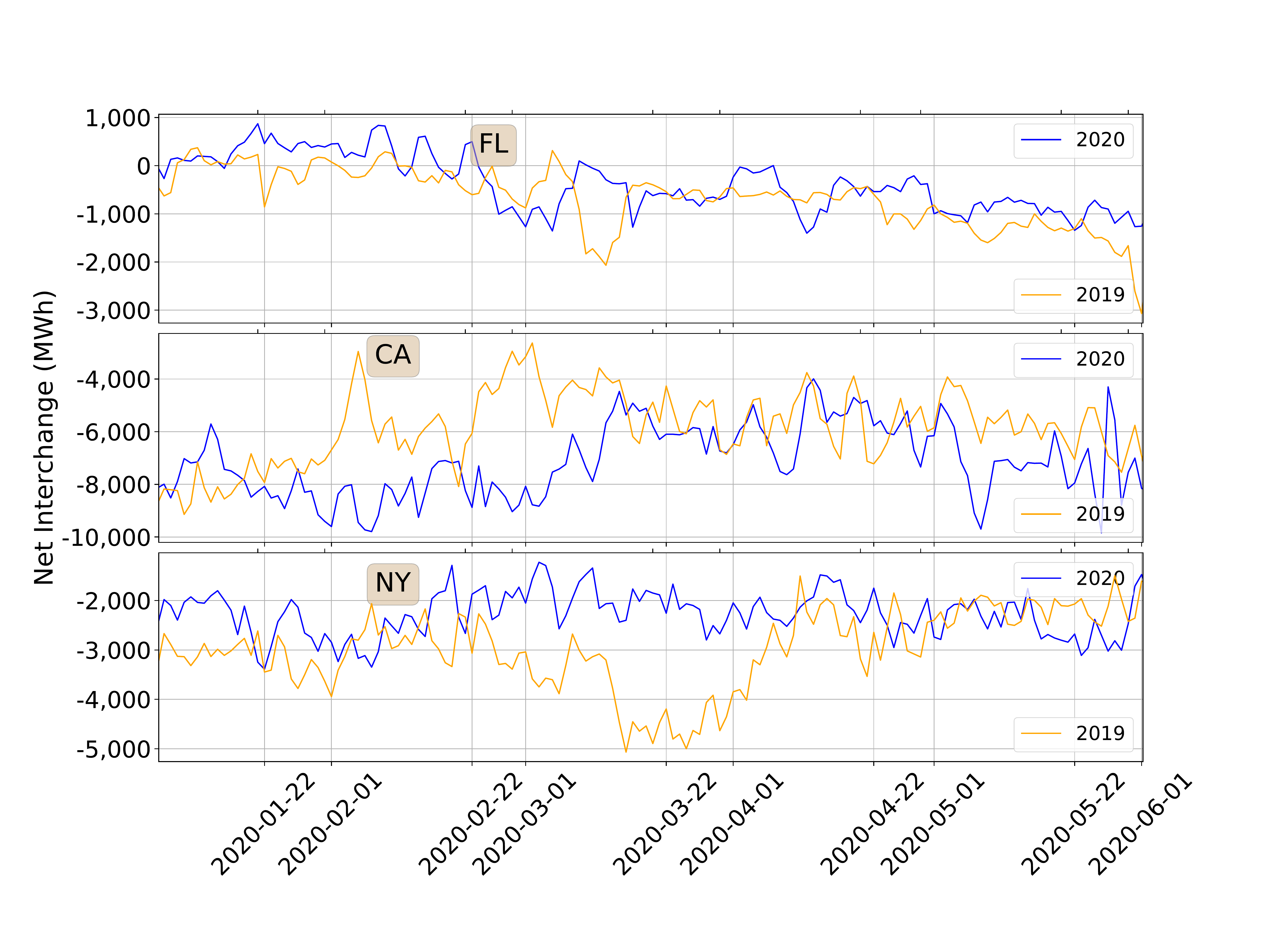}   
\caption{Mean interchange 2020 vs. 2019}
\label{fig:netInterchange}
\end{figure*}

Figure~\ref{fig:netInterchange} shows the daily average of the hourly interchange from January-May, and compares them between 2019 and 2020. There is no discernible pattern that can be attributed to the pandemic, or for that mater, anything else. The time series of hourly interchange data is highly non-stationary, with variations in both short and long time scales. We therefore do not present histograms: a much more extensive analysis will be required to reveal anything useful.

\def\mc{\mathrm{c}}
\def\mh{\mathrm{h}}
\def\alphaCool{\ensuremath{\alpha_{\mathrm{c}}}}
\def\alphaHeat{\ensuremath{\alpha_{\mathrm{h}}}}
\def\TcoolStpt{\ensuremath{T_{\mathrm{c,stpt}}}}
\def\TheatStpt{\ensuremath{T_{\mathrm{h,stpt}}}}
\def\R{\mathbb{R}}
\section{Controlling for weather}\label{sec:controlling-for-weather-method}
Methods for weather correcting energy/electricity demand data are typically based on \emph{cooling degree day} and \emph{heating degree day}. The cooling degree day of a time duration with average temperature $T$ is the the \emph{cooling degree} $T^C=\max(T-\TcoolStpt, 0)$ times the number of days that the cooling degree is positive. The \emph{cooling setpoint}  $\TcoolStpt$ is an average measure of ambient temperature above which consumers start using space cooling\footnote{Formally, the \emph{cooling degree time of an arbitrary time period $\tau$} is the area under the cooling degree curve plotted as a function of $t$. That is, $CDT[\tau] = \int_{0}^\tau T^C(t)dt$, where $T^C(t) = max(T(t)-\TcoolStpt,0)$ is the time varying cooling degree. If the integral is expressed in \degree F-hour (or \degree C-hour), we call it a cooling degree hour (CDH). Frequently it is expressed in \degree F-day (or \degree C-day), and we call it a cooling degree day (CDD). A hypothetical day that is a constant $2\degree$ F hotter than the cooling setpoint throughout the day will have a CDH of $ 48$ $\degree$F-hour, or a CDD of $2$\degree F-day. A day that is $2\degree$ F hotter than the cooling setpoint for 6 hours of the day will have a CDH of $12$ $\degree$F-hour and a CDD of $0.5$\degree F-day. A similar definition is used for heating degree hour (HDH) or heating degree day (HDD). Heating degree day (HDD) has a similar definition, except now the excess temperature is the \emph{heating temperature} $T^H(t):=max(\TheatStpt-T(t),0)$.}. Heating degree day (HDD) has a similar definition, except now the excess temperature is the \emph{heating temperature} $T^H :=\max(\TheatStpt-T, 0) $with where the \emph{heating setpoint} $\TheatStpt$  is an average measure of ambient temperature below which consumers start using space heating.

In the simplest model used for weather correction, electricity demand in a time duration with average temperature $T$, and corresponding cooling degree day $T^C$ and heating degree day $T^H$, is modeled as $d = \alpha^C T^C+ \alpha^HT^H +b$ where $b$ is the baseload, i.e. the weather-independent part of the demand, and $\alpha^C,\alpha^C$ are coefficients. To perform weather correction, the unknown constants $\alpha^H,\alpha^C,b$ are determined from \emph{pre-event} data. Then, the demand for a time interval of equal length \emph{after the event occurs} is predicted as $\hat{d} = \alpha^CT_k^C+\alpha^HT_k^H+b$, where $T_k^C,T_k^H$ are now computed from mean ambient temperature observed \emph{after the event}, with $\alpha^C, \alpha^H,b$ previously estimated. This prediction $\hat{d}$ is a counterfactual demand, the demand that should have have been observed under the post-event weather if nothing other than the weather were to change from pre-event to post event. The prediction $\hat{d}$ is referred to as the weather corrected demand\footnote{There are in fact many methods for weather correction of energy data, though they almost always use concepts of cooling and heating degree days. These methods are known to suffer from many weaknesses. An excellent summary of the difficulties with weather correction using degree days can be found in~\cite{BizEEDegreeDays:2020}.}

For greater resolution and accuracy, the baseline demand can be modeled as a time varying quantity. This is what we do in this paper, and describe next. Recall that each hour is denoted by a discrete counter $k=1,\dots$, and $d_k$ is the electricity demand during hour $k$ and $T_k$ is the mean ambient temperature during hour $k$. The cooling and heating day definitions above are applied to hourly samples, leading to a cooling degree $T^C_k = \max(T_k - \TcoolStpt,0)$ and heating degree $T^C_k = \max(\TheatStpt-T_k,0)$ at hour $k$, etc. We use the following model of hourly electricity demand, where hour $k=1$ corresponds to 01:00:00 of Monday (morning):
\begin{align}
  \label{eq:LR-2}
  d_k= \alpha^H (T^H_k)^2 + \alpha^C (T^C_k)^2 + b_{k}, \quad k=1,\dots,168 (=24 \times 7)
\end{align}
Justification for the quadratic terms is that the data we used (described in the next section) showed such a relationship---when electric energy demand was plotted against cooling and heating degrees. The quantity $b_k$ is the hourly baseline demand, which is independent of weather. The index $k$ in~\eqref{eq:LR-2} only goes up to 7 days: we have \emph{assumed that consumer behavior that determines weather-independent electricity consumption does not change from one week to another.} 


If we only have one week's worth of hourly data, there are $2+168=170$ variables in this model, with $168$ measurements so the model is highly under-determined. Almost any week's data can be explained by such a model extremely well, but its predictive power cannot be assessed since there is no data left to be used for out-of-sample testing.  We can use additional data from other weeks which will increase the number of equations of the form~\eqref{eq:LR-2},  while the  number of parameters to fit remain the same since we have assumed that the baseload $b_k$ does not change from week to week. The 170 unknown parameters can then be estimated from the data, for instance, by ordinary least squares.

Suppose we have $m$ weeks of data. With hourly data collection, the sample index $k$ goes from $1$ to $168 \times m$. To be able to describe the model and the data in standard regression form, we write it as
\begin{align}
  \label{eq:LR-3}
  d_k= \alpha^H (T^H_k)^2 + \alpha^C (T^C_k)^2 + b_{k'}, \quad k=1,\dots,m\times 168
\end{align}
where $  k' = k \mod 168$. The model~\eqref{eq:LR-2} can be written compactly as $ d_k = \phi_k \theta$,
where the unknown parameter vector $\theta$ is
\begin{align}
  \label{eq:3}
  \theta = [\alpha^H,\alpha^C,b_1,\dots,b_{168}]^T \in \R^{170}
\end{align}
and the regressor is the row vector
\begin{align}
  \label{eq:5}
  \phi_k=
  \begin{bmatrix}
    (T^H_k)^2 & (T^C_k)^2 & 0 & 0 & \dots & 0 &1 & 0 & \dots 
  \end{bmatrix}
\end{align}
in which the $1$ appears on the $2+(k \mod 168)$-th column position.  The corresponding linear equation becomes
\begin{align}
  \label{eq:4}
  d = \Phi \theta
\end{align}
where $d = [d_1,d_2,\dots,d_n]^T \in \R^n$,  with $n = 168 \times m$ denoting the total number of hourly time samples, and $\Phi = [\phi_1^T,\phi_2^T,\dots,\phi_n^T]^T \in \R^{n \times 170}$. The ordinary least square (OLS) estimate of the unknown parameter $\theta$ is the solution to the normal equation\footnote{Technical requirements such as full column rank of $\Phi$ are satisfied by the data we applied the method to. Ideally one should pose the model training problem as a constrained optimization problem with the constraint that all the unknown parameters are non negative, but OLS never returned a negative estimate for the data we used, meaning that if a quadratic cost were used the solution will be the same.}:
\begin{align}
  \label{eq:6}
(\Phi^T\Phi) \theta = \Phi^Td \Rightarrow \hat{\theta} =   (\Phi^T\Phi)^{-1}\Phi^T d.
\end{align}
The weather correction method with this model is the same as in the simpler model. Suppose the model is trained with data from year 2019, and denote by $\hat{\theta}^{(2019)}$ the parameter estimate from the training. The weather corrected demand at hour $k$ during 2020 is the demand predicted by the model for weather during the same hour of week in 2020:
\begin{align}
  \hat{d}^{(2020/2019)}_k = \phi_k^{2020} \hat{\theta}^{(2019)} ,
\end{align}
where $ \phi_k^{2020}$ is the regressor for that hour during 2020, which is constructed from temperature measurements at hour $k$ in 2020 and time of week that $k$ corresponds to. The quantity $\hat{d}^{(2020/2019)}_k$ is counterfactual, it is the demand that should have been observed in 2020 had nothing other than the weather changed from 2019 to 2020.

  \begin{figure*}[ht]
  \centering
  \includegraphics[width=0.9\textwidth]{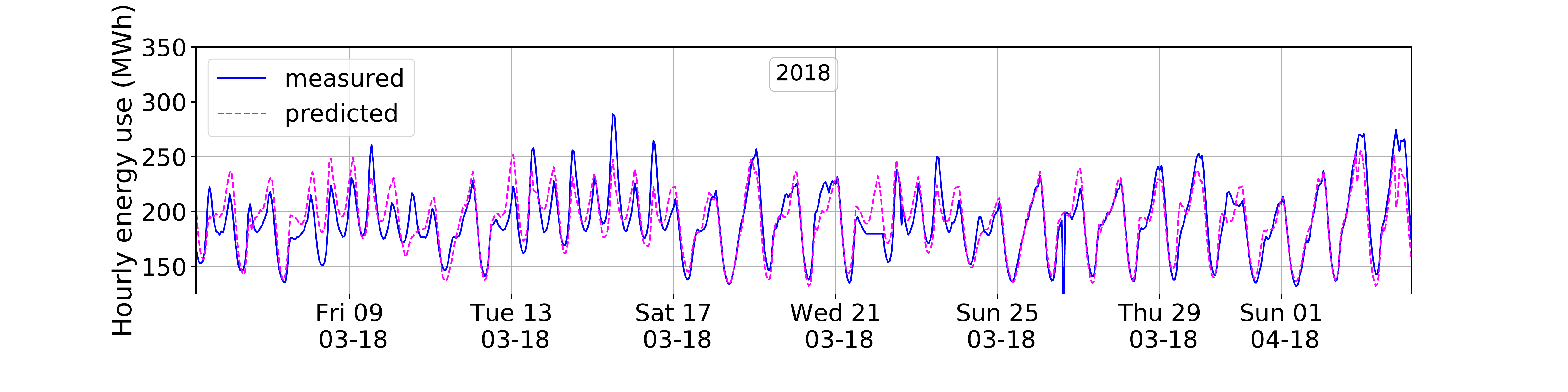}\\
  \includegraphics[width=0.9\textwidth]{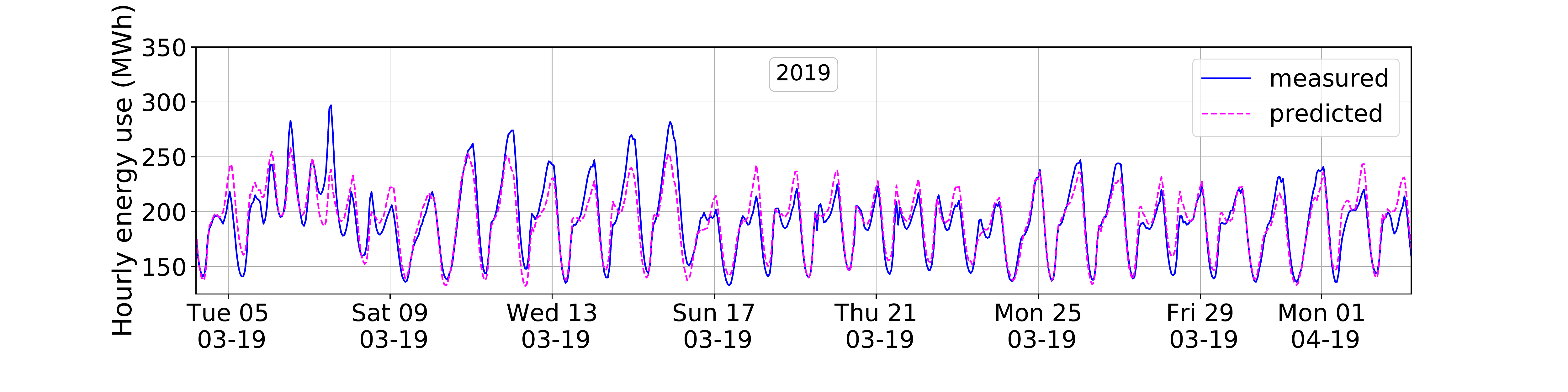}\\
  \includegraphics[width=0.9\textwidth]{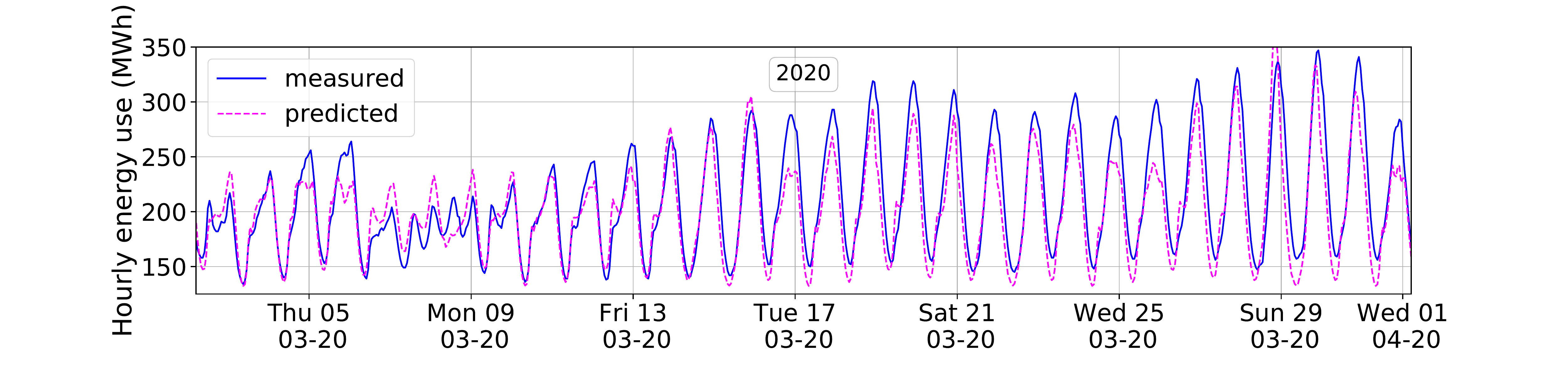}
  \caption{Hourly demand prediction for 3 distinct years by the model~\eqref{eq:LR-2} trained with 2019 March data. The x-axis starts on the first Monday of March for each year. }
  \label{fig:trainingTestingPerformance}
\end{figure*}

An example of the model's in-sample and out-of-sample prediction is shown in Figure~\ref{fig:trainingTestingPerformance}. The data used is described in Section~\ref{sec:GRU}. The model was trained with 2019  March data, so the predictions  for 2020 and 2018 are  out-of-sample predictions. The mean and standard deviation of the fitting error $e_k = (d_k^{2019} - \hat{d}_k^{(2019/2019)})/d_k^{2019}$ are $-1.4\%$ and $7.7\%$, respectively. The $R^2$ value of the fit is $72\%$. 

\subsection{Application to GRU (Gainesville Regional Utilities)}\label{sec:GRU}
The immediate difficulty in applying the weather correction method described above is the ambiguity in defining the ambient temperature $T_k$. One can perhaps take a weighted average over many weather stations, with measurement of a station weighted by the number of consumers around that station etc., but this method introduces more design choices that affect the model's prediction. To avoid this, we apply the weather correction method to Gainesville Regional Utilities (GRU). GRU's service area is confined to the city of Gainesville, FL, and a small area nearby. Apart from being an electric utility, it is also a balancing authority (BA). Because GRU is a BA, hourly data from GRU is available from EIA.

The model~\eqref{eq:LR-2} was trained with hourly electricity data for GRU and weather data for Gainesville, for March 2, 2019, to March 31, 2019\footnote{EIA's data was missing for several hours on March 1, 2019.}. It was then used to predict the energy consumption for March and April 2020 by using weather  data for that time period. Temperature measurements from the University of Florida (UF) weather station, available through alachua.weatherstem.com was used, since the UF campus is in the middle of GRU territory. We chose the values of the setpoints $\TheatStpt,\TcoolStpt$ to be $64,72$\degree\ F, respectively, based on an exhaustive search that led to the smallest fitting errors. 

\begin{figure*}[htb]
  \centering
  \includegraphics[width=0.95\textwidth]{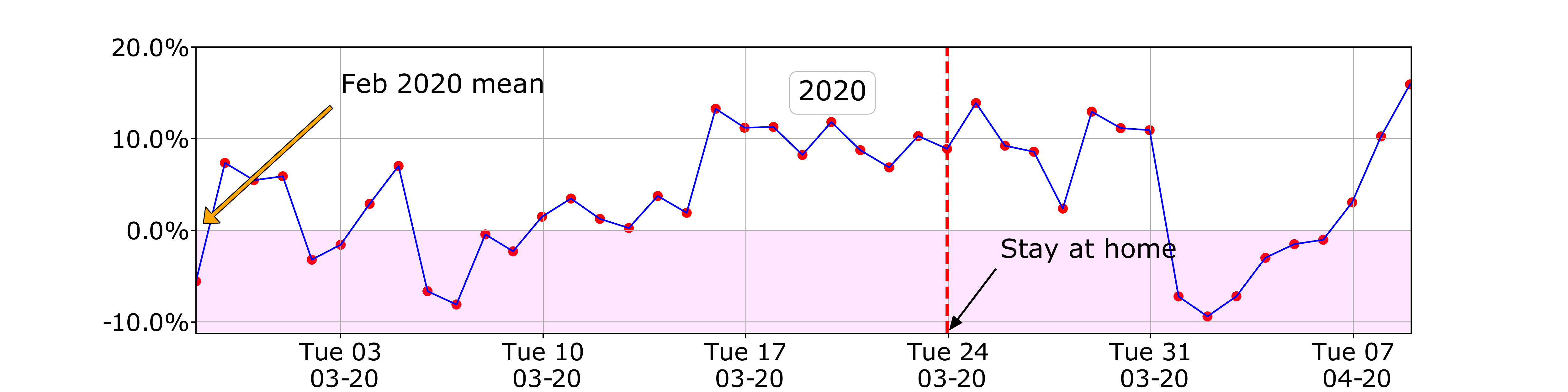}\\
  \caption{Weather corrected daily energy demand in 2020 for GRU: difference between 2020 demand and its predicted value by the model~\eqref{eq:LR-2}. The stay-at-home order was issued by the Alachua county to take effect on March 24, 2020.}
  \label{fig:changeInDemand-2020}
\end{figure*}

Figure~\ref{fig:changeInDemand-2020} (top) shows the change in daily electricity demand during March-April of 2020 from the counterfactual demand with 2020 weather, i.e., what should have been the demand had there been no change in baseload demand from the same time 2019. More precisely, it shows the difference
\begin{align}\label{eq:e}
e_t^{2020/2019} := d_t^{2020} - \hat{d}_t^{(2020/2019)}  
\end{align}
 in the daily electricity use $d_t$  as a percentage of average daily energy use in February 2020; the index $t$ increments by 1 every day. 

\begin{figure*}[htb]
  \centering
  \includegraphics[width=0.85\textwidth]{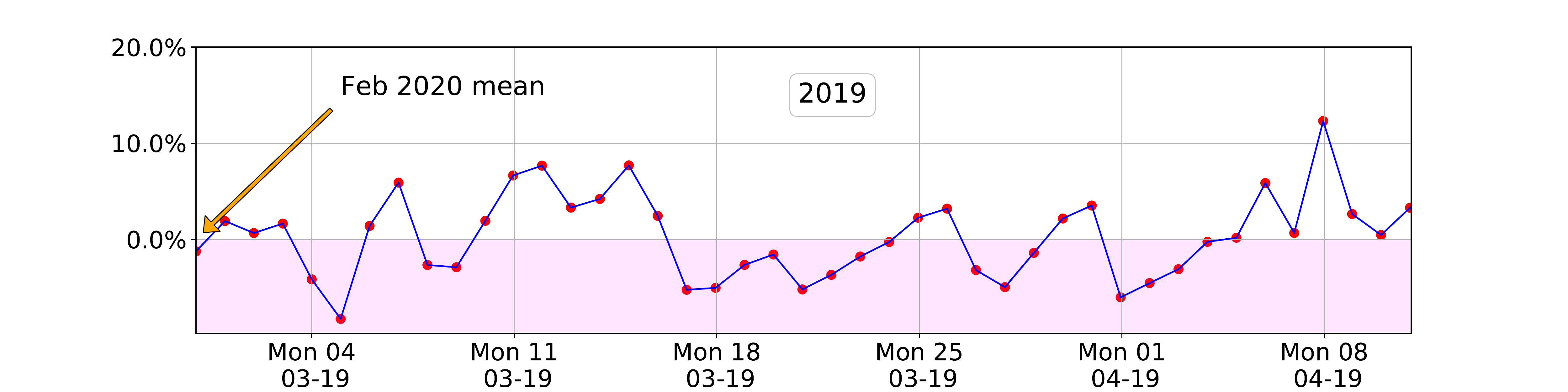}
  \includegraphics[width=0.85\textwidth]{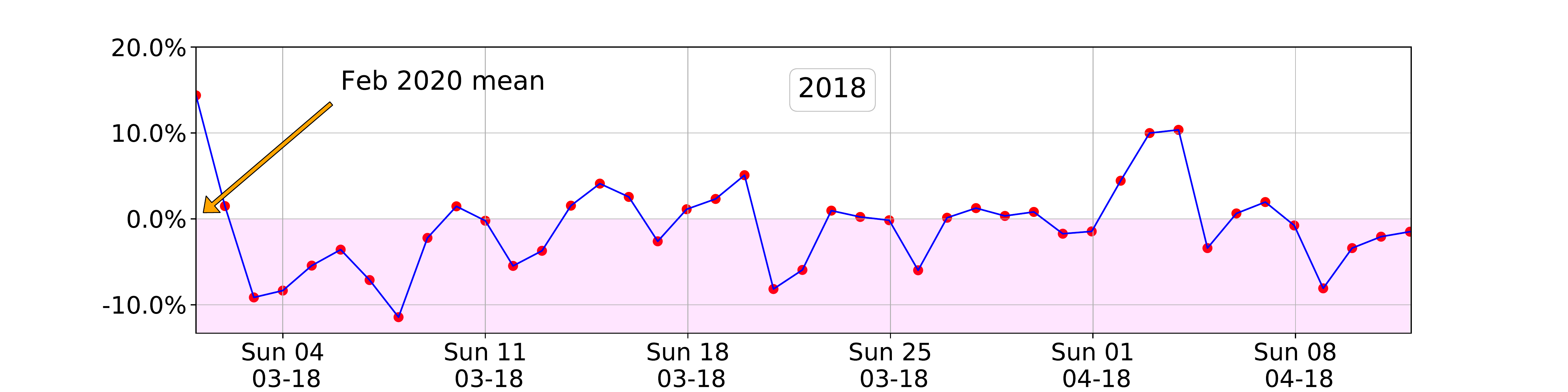}
  \caption{Change in weather corrected daily energy demand for GRU during March-April for two pre-pandemic datasets. The model is trained with 2019 data, so one of them shows the in-sample prediction error.}
  \label{fig:changeInDemand-2018-2019}
\end{figure*}

It appears there is an increase of about 10\% in daily electricity demand around the stay-at-home order issued by the Alachua city that GRU territory is within. This raises a number of questions, which we address next.
\begin{enumerate}
\item {\bf Is it really a 10\% increase?} The mean and standard deviation of the prediction error $e_t$ for the training data is $0.6\%$ and $3.9\%$ and similar for the out-sample testing data from 2018; see Figure~\ref{fig:changeInDemand-2018-2019}. Since the mean is so small, we assume the model's predictions are unbiased, and use 0 as the mean error for out of sample predictions. Thus, we can conclude that there is a 95\% probability that the change in daily demand is between $2\%$ and $16\%$ of the February 2020 mean\footnote{The 95\% confidence interval is $\pm 2\sigma$ away from the mean, and 99\% confidence interval is $\pm 3\sigma$, assuming the errors are Gaussian. The residual of the trained model show Gaussian-like distribution (figure not shown).}. There is a 99\% probability the increase is between $-2\%$ and $22\%$. Thus, while there is a \emph{likely} increase in demand, quantifying the increase with higher confidence will require additional modeling and analysis.
\item {\bf Why not ``improve'' the model?} The model only resolves baseload demand within a week, but ignores seasonality over longer time scales. To avoid overfitting, we must use a sufficient number of weeks of data for training,  so that $n=168\times m \gg 170$. At the same time, we cannot use too many weeks of data, since that data will surely violate the assumption that weather-independent baseload does not change from week to week. One can allow the baseload to vary across weeks or months, introducing additional free parameters in the model. One can also introduce more free parameters for holidays and special events. But doing so will soon lead to overfitting: the model will have nearly as many free parameters as the number of measurements used to estimate them. There will be little data left to test for out-of-sample prediction accuracy of the model. For the model to be trustworthy, it is not enough for the model to fit the training data accurately. The out-of-sample prediction accuracy of the model also needs to be high. Only then, we can have confidence on the underlying assumptions of the model.  Since the model is trained for March data, its validity reduces as we go further away from March.Therefore,we do not use it for May and beyond.  
\item {\bf Why did the demand increase before the stay-at-home order?} Assuming the model's prediction of an increase in electricity demand is correct, why did the increase occur even before the order? In other words, can we blame the pandemic for the increase in demand? We believe so. Although GRU does not serve the University of Florida campus, the majority of Gainesville residents and, thus, GRU customers are directly or indirectly related to the university. In an email on March 11, 2020, the University declared that all classes will be moved online by Monday, March 16, and recommended all students to return to their homes. It is thus very likely that many GRU customers started to work from home by March 16, 2020, and did not wait until March 24, when Alachua county issued its stay-at-home order\footnote{This is certainty true for the authors of this paper.}. This may be the reason for the increase seen from March 17 onwards in Figure~\ref{fig:changeInDemand-2020}. \\
Although we can only speculate for the reasons of the increase in electricity consumption due to the pandemic in GRU, it is possible that the large residential consumer base of GRU, coupled with the lack of large industries in its territory, plays a role. If that is the case, this is an example in which reduction of electricity consumption in commercial buildings after the pandemic is more than offset by the increase in consumption in residences due to more people staying and working from home. 
\end{enumerate}

\section{Summary and conclusion}
Unlike previous analyses that were done soon after the pandemic started, our analysis benefits from examining data over a longer period of time. We found that contrary to the observations in the prior work, none of these three regions analyzed (New York, California, and Florida) showed a clear reduction in the demand that can be ascribed to the pandemic.

In both electricity demand and variables such as peak demand that can indicate stress on the power grid, we observed large variability from one region to another. This is consistent with EIA's recent analysis on electricity demand across the USA. In addition, while some of the indicators showed a change in their statistics around the time stay-at-home orders were issued (compared to their values from the same period in 2019), they seemed to revert back to their pre-pandemic or 2019 values by May 2020. Interestingly, some indicators of stress indicated an increase in stress while others indicated the opposite. If these trends are correct, that would mean the change in consumers' behavior due to the pandemic mitigation efforts---in so far as it affects the power grid---was temporary.


Almost all earlier studies reported a reduction in electricity demand coincident with COVID-19 mitigation efforts. Since the pandemic is reducing economic activity, a reduction in energy demand is expected. When it comes to electricity, such a conclusion is less obvious. According to the EIA, buildings consume 75\% of the electricity in the US~\cite{eiaElectricityExplained:2020}. While electricity consumption in commercial buildings (offices, schools, retail stores, restaurants etc.) may have reduced, it is possible that such a reduction is offset by an increase in residential building electricity consumption since people are staying and working from home. Resolving this question requires that we estimate the change due to other concomitant factors, especially weather. 

The weather correction exercise for a small balancing authority in Florida indicated that the pandemic led to an increase in electricity demand there. Our speculation for the increase is that the more people staying home is increasing residential electricity demand, and the reduction in commercial demand is not enough to offset it. Whether it is going to continue, or whether such a trend is likely in larger regions, will require further work.

The analysis presented here is a first step. It needs to be continued and refined as the pandemic---and the mitigation efforts to contain it---continues to evolve, to understand the impact on consumer behavior and electricity sector.

The weather correction exercise also indicated that the inherent prediction error of the model is only marginally smaller than the changes predicted by the model. This makes providing quantitative estimates of the weather-corrected change challenging. Lack of careful evaluation of a model's predictive power can be dangerous: a modeler can be swayed by confirmation bias. 

\section*{Acknowledgement}
The authors thank EIA for making available the Hourly Electric Grid Monitor (\url{https://tinyurl.com/ybl2s39m}), and the weatherSTEM portal of Alachua county (\url{http://alachua.weatherstem.com/}). We also wish to thank the volunteer developers of the Python programming language and its many modules which make data analysis and visualization almost as simple as thinking about them.

\bibliographystyle{ieeetran}
\bibliography{refs}

\begin{IEEEbiography}[{\includegraphics[width=1in,height=1.25in,clip,keepaspectratio]{agdas.jpeg}}]{Duzgun Agdas} is a Senior Lecturer of Construction Engineering and Management at the Queensland University of Technology.   His research interests are infrastructure sustainability and disaster resilience.  He conducts research at QUT and with international collaborators of engineers and data scientists to find data driven solutions to grand challenges of built environment. Further information about his research projects,  publications and supervision can be found on the following webpage:
https://staff.qut.edu.au/staff/duzgun.agdas .
\end{IEEEbiography}
\begin{IEEEbiography}[{\includegraphics[width=1in,height=1.25in,clip,keepaspectratio]{Barooah.jpg}}]{Prabir Barooah} is a Professor of Mechanical and Aerospace Engineering at the University of Florida. His expertise is on modeling and control of complex engineering systems. His research focus is on developing algorithms and systems for smart resilient power grids and sustainable buildings. He directs the Distributed Control and Energy Systems (DICE) Lab at the University of Florida. Further information about his research can be found on lab webpage \url{https://dicelab.mae.ufl.edu}, his Google Scholar page,  and in \url{www.prabirbarooah.com}.
\end{IEEEbiography}

\end{document}